%% file: main.tex
\documentclass[manuscript]{acmart}

\AtBeginDocument{%
  }

\setcopyright{acmlicensed}
\copyrightyear{2025}
\acmYear{2025}
\acmDOI{XXXXXXX.XXXXXXX}

\acmISBN{978-1-4503-XXXX-X/18/06}

\usepackage{physics}
\usepackage{xcolor}
\usepackage[linesnumbered,ruled,vlined]{algorithm2e}
\usepackage{algpseudocode}
\usepackage{tikz}
\usepackage{siunitx}
\usepackage{ulem}
\usepackage{graphicx}
\usepackage{multirow}
\usepackage{hyperref}
\newtheorem{definition}{Definition}
\usepackage{enumitem}
\SetKwInput{KwInput}{Input}                
\SetKwInput{KwOutput}{Output}              
\definecolor{teal}{rgb}{0.0, 0.0, 0.0}
\usepackage{xcolor}
\newcommand{\blue}[1]{\textcolor{black}{#1}}
\begin{document}

\title{Efficient Compilation for Shuttling Trapped-Ion Machines via the Position Graph Architectural Abstraction}
\author{Bao G. Bach}
\affiliation{%
  \institution{Quantum Science and Engineering, University of Delaware}
  \city{Newark}
  \state{Delaware}
  \country{USA}
}
\email{baobach@udel.edu}

\author{Ilya Safro}
\affiliation{%
  \institution{Computer and Information Sciences, University of Delaware}
  \city{Newark}
  \state{Delaware}
  \country{USA}}
\email{isafro@udel.edu}

\author{Ed Younis}
\affiliation{%
  \institution{Computational Research Division, Lawrence Berkeley National
Laboratory}
  \city{Berkley}
  \city{California}
  \state{Delaware}
  \country{USA}
}


\begin{abstract}
    With the growth of quantum platforms for gate-based quantum computation, compilation holds a crucial role in deciding the success of the implementation. While there has been rich research in compilation techniques for the superconducting-qubit regime. The trapped-ion architectures, currently leading in robust quantum computations for their reliable operations, still lack competitive compilation strategies. This work introduces a unifying hardware abstraction, the ``position graph'', representing various hardware architectures. With this abstraction, we model trapped-ion Quantum Charge-Coupled Device (QCCD) architectures, enabling high-quality, scalable compilation methods. Specifically, we propose scheduling algorithms called SHuttling-Aware PERmutative (SHAPER) and SHuttling-AWare (SHAW) heuristic searches to tackle the complex constraints and dynamics of trapped-ion machines, with the cooperation of state-of-the-art permutation-aware mapping. These approaches generate executable circuits and native instructions that respect the physical constraints of shuttling-based architectures. We evaluate proposed algorithms across theorized and real architectures using the position graph framework. For completeness, we also introduce a linear program of trapped-ion scheduling that yields the optimal schedules, enabling a direct comparison with our heuristics. Our algorithm can successfully compile programs for extreme architectures where priori algorithms fail. When the baseline does complete, our produced schedules are \color{teal} $1.45$ times \color{black} faster on average, with best-case speedups up \color{teal}to $4$ times faster\color{black}. \\ 
    {\bf Reproducibility:} source code and computational results are available at $[$will be added here upon  acceptance$]$.
\end{abstract}




\keywords{Quantum compilation, Trapped-ion architecture, Quantum charge-coupled devices, Quantum circuit mapping}


\maketitle


\input{isca2024/content/1introduction}
\input{isca2024/content/6related_works}
\input{isca2024/content/2background}
\input{isca2024/content/2QCCD}

\input{isca2024/content/3position_graph}

\input{isca2024/content/4QCCD_mapping}

\input{isca2024/content/5evaluation}

\input{isca2024/content/7discussion}
\input{isca2024/content/8conclusion}






\begin{acks}
B.B and I.S are supported with funding from the Defense Advanced Research Projects Agency (DARPA) under the ONISQ program. E.Y. is supported by the U.S. Department of Energy, Office of Science, Office of Advanced Scientific Computing Research under Contract No. DE-AC05-00OR22725, through the Accelerated Research in Quantum Computing Program MACH-Q
\end{acks}

\bibliographystyle{ACM-Reference-Format}
\bibliography{refs}
\newpage









\appendix
\input{isca2024/content/9appendix}
\end{document}

%% file: isca2024/content/1introduction.tex
\section{Introduction}
\label{sec:Intro}
Quantum computing is rapidly evolving into a transformative technology with significant potential in areas such as finance \cite{herman2023quantum}, simulations in chemistry \cite{cao2019quantum}, combinatorial optimization \cite{shaydulin2019hybrid}, and machine learning \cite{biamonte2017quantum}, to mention just a few. We are currently in the \emph{noisy intermediate-scale quantum} (NISQ) computing era, characterized by quantum devices with tens to thousands of noisy qubits, short coherence times, and limited connectivity. One of the key challenges in this era is the efficient compilation of quantum circuits to address the specific limitations and architectural constraints of various quantum hardware platforms.

\begin{figure}
    \centering
    \includegraphics[width=.6\linewidth]{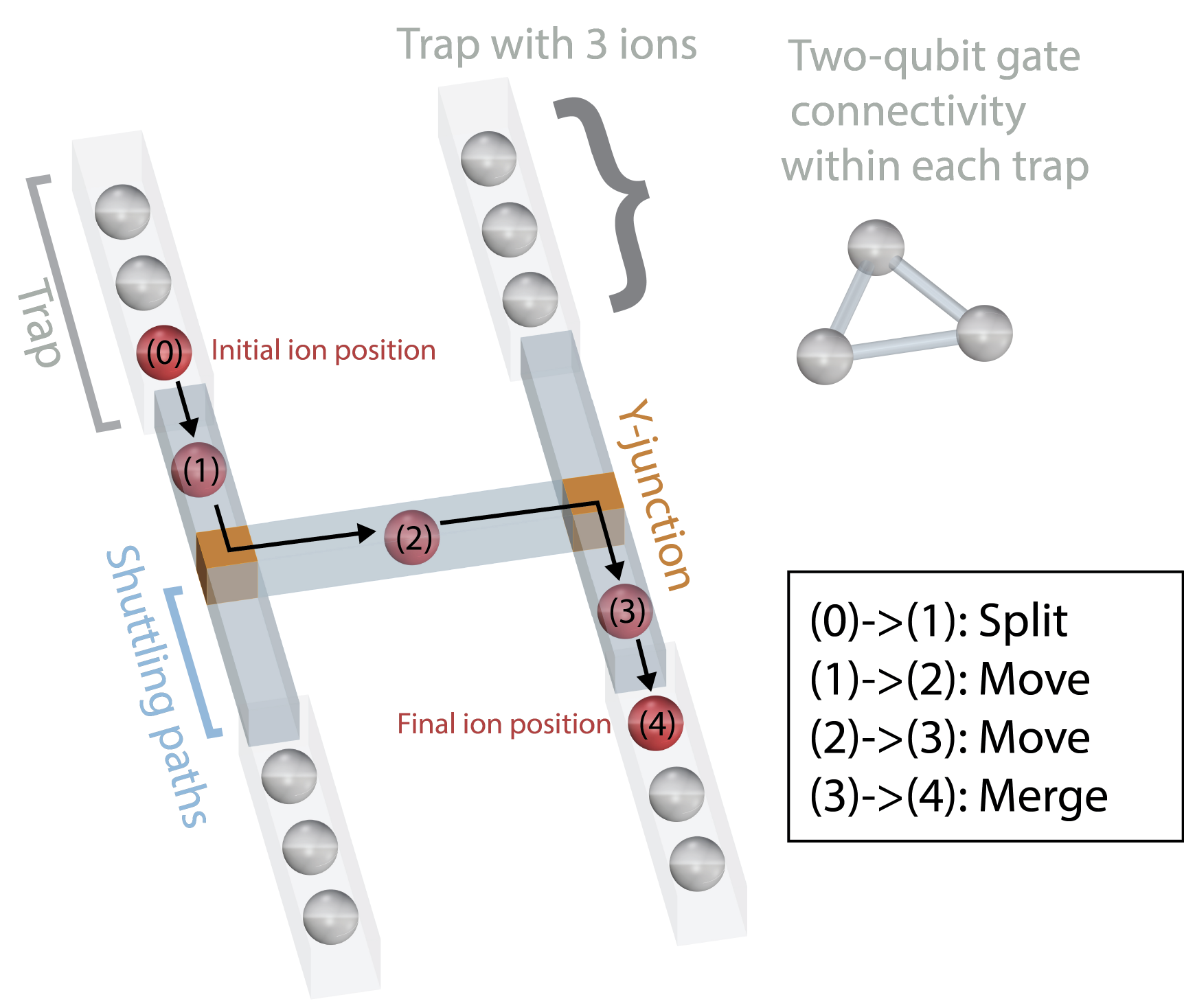}
    \caption{The \textbf{H}-type \cite{murali2020architecting, kielpinski2002architecture} QCCD-based trapped-ion architecture. Here, the ion is moved from one trap to another trap using 3 shuttling operations. Move $(0 \rightarrow 1)$ denotes a split operation to split the ion from the trap. Move $(1 \rightarrow 2)$ and $(2 \rightarrow 3)$ denotes a move through a Y-junction (T-junction) into the segment space. Finally, move $(3 \rightarrow 4)$ shows the merge operation of the ion into the new trap.}
    \label{fig:H_QCCD}
\end{figure}

Among the various quantum computing architectures, \emph{trapped-ion} (TI) devices stand out as one of the most promising options currently, arguably offering the most robust computation ~\cite{decross2024computational, shapira2018robust, weber2024robust}. These devices confine atomic ions within traps and utilize two of the atomic states to represent quantum information. Operations are executed using precisely calibrated lasers that are directed at the trap. The unique physics of trapped-ion architectures allows for high-fidelity operations and long coherence times at the cost of long execution times. For example, trapped-ion operations are measured in microseconds, whereas superconducting machine instructions are measured in nanoseconds~\cite{linke2017experimental}. To make this matter worse, in the current state-of-the-art architecture scheme called \emph{quantum charge-coupled device} (QCCD), the trapped ions must be physically shuttled around the chip during program execution to create connections. Figure \ref{fig:H_QCCD} demonstrates an example of QCCD architecture and the use of shuttling operations to move ions around. These slow shuttling operations comprise most of the final program's runtime.
\color{teal} Following the gate fidelity model used in \cite{murali2020architecting}, we associate operations runtime directly to accumulated heating, which affects the reliability \cite{blakestad2011near}. This fidelity model is discussed in details in Section \ref{sec:exp_setup}. 
This motivates establishing high-quality compilation practices for TI devices, which is absolutely crucial as the runtime overhead can be significant. \color{black} These points underscore the importance of high-quality compilation practices for trapped-ion devices.

The usual compilation paradigm employed for TI systems usually involves a two-layer approach \cite{kreppel2023quantum, schoenberger2024shuttling}. First, the logical layer abstracts away the device's unique physical constraints by assuming all qubits are connected in a simple all-to-all model. Here, compilers focus on optimizing a program in a machine-agnostic way. The second compilation step is then to resolve the physical constraints given the logical program. This involves breaking the logical operations up with shuttling instructions to create a schedule of shuttling instructions. While this all-to-all assumption simplifies some aspects of the compilation process, it paradoxically complicates the ability to accurately and effectively account for the underlying hardware's unique constraints. \emph{We argue that this two-layer approach is not well-suited for producing efficient shuttling schedules that minimize execution time and, as a result, maximize reliability.} This is because the hardware-agnostic stage does not consider shuttling and may introduce a seemingly great optimization only to make it worse during shuttling. For example, during the circuit optimization, gates that act on the far-distance and nearby ions will be indistinguishable. This can lead to accumulated far-distance ion coupling, which introduces an underlying overhead. 

\begin{figure}
    \centering
    \includegraphics[width=\linewidth]{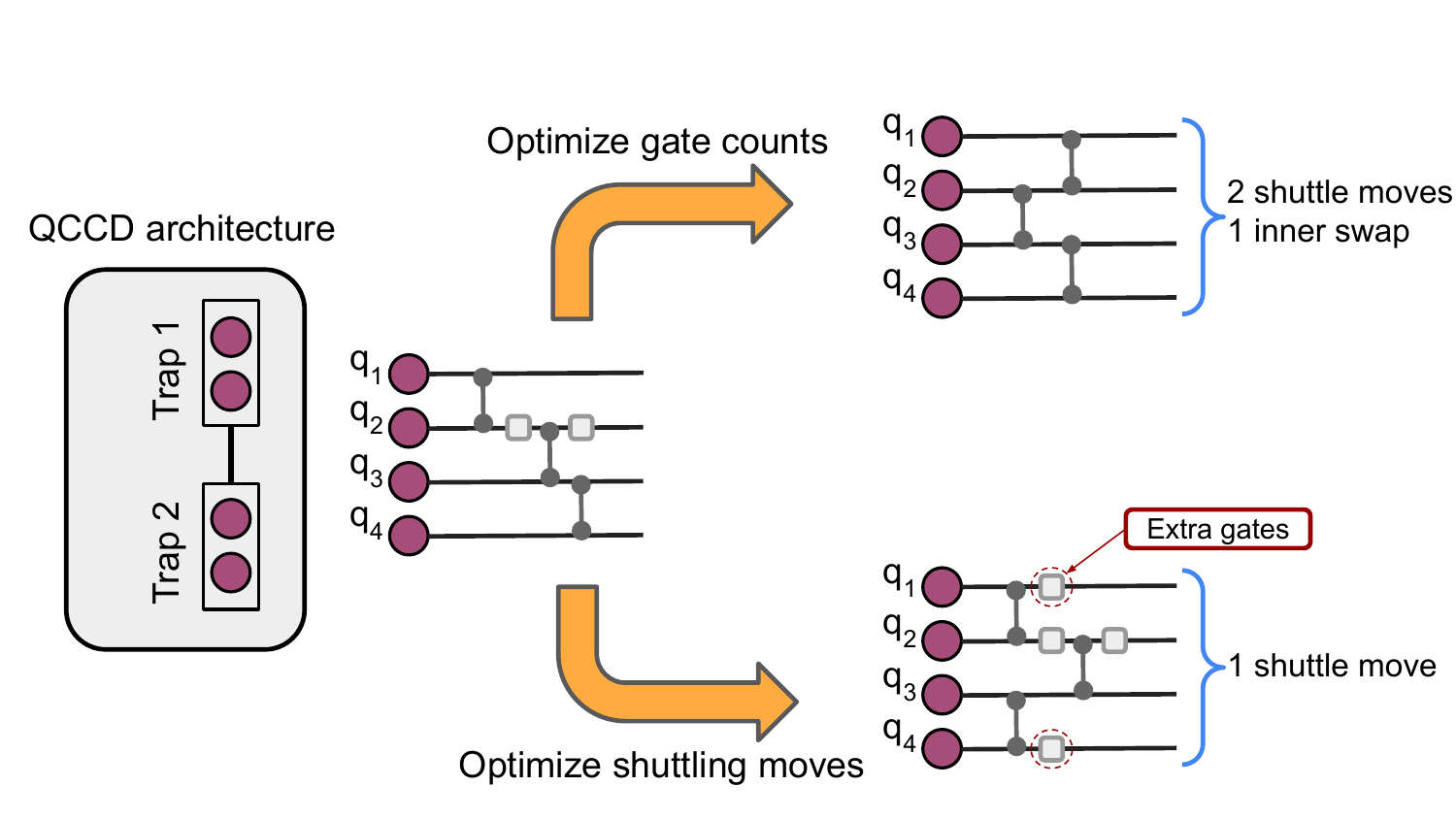}
    \caption{An example of circuit compilation using position graph abstraction, where the number of shuttling moves is optimized; and using common abstraction, where the number of gate counts is optimized. This trade-off is important for the QCCD-based TI cases, as the cost of shuttling is more expensive than gate execution.}
    \label{fig:concrete_ex}
\end{figure}

To give a concrete example of this problem, we show in Fig. \ref{fig:concrete_ex} a possible scenario when considering the all-to-all assumption, the optimizer moves the gate around to eliminate the single-qubit gates, but at the cost of introducing shuttling moves. 
QCCD trades qubit-count scaling for a connectivity challenge: all-to-all holds only within a trap. Treating the device as fully all-to-all encourages compilers to optimize gates first and only then ‘repair’ shuttling—often inflating movement. With our novel position graph abstraction introduced below, we optimize the circuit concerning the abstracted architecture and come up with operations that minimize the shuttling operations. 

Consequently, there is a pressing need to break away from the limitations imposed by the two-layer approach and formulate a more meaningful abstraction of the constraints and shuttling operations inherent to QCCD architectures. However, this presents several challenges as the complex constraints yielded by the physical architectures cannot be captured on the well-known coupling graph abstraction \cite{siraichi2018qubit}, and there is a lack of physical abstraction for the shuttling-based compilation schedule. Introducing an effective abstraction that can lead to more efficient compilation schemes for QCCD-based shuttling architecture is extremely important. This work attempts to address the following fundamental question:
\textit{How do we break the assumption of all-to-all qubit connectivity and its resulting two-stage approach to TI compilation, and instead derive a meaningful abstraction for the constraints and shuttling operations of QCCD architectures for the compilation process?}
Compilers for superconducting devices abstract the entire architecture to a coupling graph, which reflects the physical connectivity of the qubits \cite{siraichi2018qubit}. That is, multi-qubit gates can only be applied to qubits connected in the graph. All other machine-level details are abstracted away, and as a result, numerous algorithms have been developed for compiling these graphs. 

In this paper, we introduce the \textit{position graph}, an extension to the coupling graph abstraction, as the answer to the posed question and a unifying abstraction between different architecture types. This abstraction provides a framework for encoding hardware constraints while enabling the same superconducting algorithms to be trivially adapted for the QCCD case. The position graph is a thin middle-layer abstraction in the quantum compilation stack, allowing for broader applicability of compilers. We argue that this abstraction naturally implies physical restrictions of the shuttling operations and QCCD system, as will be discussed in Section \ref{sec:QCCD}. To further support the introduction of the position graph, we relate the concept to the well-known theoretical foundation of token reconfiguration and cooperative pathfinding by showing the similarity between the problems through the use of the position graph. Moreover, the initial mapping of \color{teal}program qubits \color{black} to physical ions and the transitions of physical ions between traps using shuttling operations can be generalized through this abstraction. 

To demonstrate the efficacy of our proposed abstraction, we successfully designed a mapping and routing algorithm for the position graph representation. Our algorithm is inspired by the state-of-the-art mappings for superconducting qubits. Our findings indicate that this adaptation competes effectively with existing shuttling algorithms and enhances them by handling a wider range of cases. In particular, our approach is compatible with any gate set and offers solutions for congestion issues without oversimplifying the hardware constraints. While other approaches either \emph{fail to resolve the deadlocks} created by shuttling operations or try to simplify the constraints, making them unusable in practice \cite{murali2020architecting, schoenberger2024shuttling, kreppel2023quantum}, our proposed \emph{SHuttling-Aware PERmutative search algorithm} (SHAPER) resolves scenarios where the state-of-the-art fails. Specifically, given a QCCD-based TI system with $m$ traps, each having a maximum ion capacity of $n$, our \textit{position graph}-based method manages to compile any quantum circuit with $m n$ qubits. In contrast, in the best case scenarios, state-of-the-art available tools can only compile circuits with $(m-1)  n$ qubits \cite{murali2020architecting}.\\
In summary, our contributions are: 
\begin{itemize}[leftmargin=1em,topsep=0cm]
    \item We introduce the \textit{position graph} abstraction to bridge the gap between the rich literature in circuit mapping problems and the shuttling problem in TI systems. To the best of our knowledge, this is the first attempt to narrow the difference between \color{teal}\sout{different quantum architecture abstractions} QCCD-based TI and superconducting architecture abstraction \color{black}, which has the potential to unify compilation processes.
    \item We propose the SHuttling-Aware PERmutative search algorithm (SHAPER) and SHuttling-AWare heuristic search (SHAW) designed using the \textit{position graph} by porting a state-of-the-art mapping algorithm from superconducting systems. The algorithms are benchmarked across various quantum circuits and architectures, demonstrating successful results across all configurations \emph{where the state-of-the-art methods fail}. For executable configurations, our approach consistently delivers superior performance in most cases.
    \item We demonstrate the first shuttling-based algorithm that can resolve congestion and deadlocks as well as fully utilize two-dimensional QCCD-based architectures. We further link the rich, theoretically based token reconfiguration and cooperative pathfinding to the shuttling problem and hinge on an open future direction for adaptation.
    \item We introduce a mixed integer linear programming formulation for the trapped-ion scheduling problem to achieve high-quality but slower solutions and to serve as a comparison baseline. Using this formulation, we solve the problem on 8-qubit circuits, and compare with our proposed heuristic SHAPER, SHAW, and the available tools QCCDSim. The formulation can be used for larger circuits as well, provided more significant computational resources.
\end{itemize}
The rest of the paper is organized as follows. Related work is presented in Section \ref{sec:related_works}. Section \ref{sec:QIC} gives a brief introduction to quantum information, quantum compilation, and a permutation-aware mapping strategy. Section \ref{sec:QCCD} introduces the Quantum Charge-coupled device (QCCD) system, the shuttling operations, and their constraints. Sections \ref{sec:position_graph} and \ref{sec:our_algo} introduce the \textit{position graph} abstraction and mapping algorithm that is based on it, which are the main novel points of the paper. We present the experimental results in Section \ref{sec:evaluation}.  We conclude our paper with a discussion in Section \ref{sec:discussion}.

%% file: isca2024/content/6related_works.tex
\section{Related work}
\label{sec:related_works}
There has been significant research in the compilation of superconducting qubit systems. In \cite{siraichi2018qubit, li2019tackling, liu2023tackling, zhu2020exact}, the authors were concerned with finding the best mapping based on the original circuit. In \cite{lin2013ftqls, davis2020towards, younis2021qfast}, the goal was to find the best-synthesized circuit concerning the hardware connectivity graph by performing synthesis. Different techniques (e.g., machine learning) for finding the best-synthesized circuit for the connectivity graph are well-studied \cite{moro2021quantum, duong2022quantum, zhang2020topological}. A more detailed review of quantum compilation and synthesis of superconducting-based systems can be found in \cite{ge2024quantum}.

In the opposite situation, the QCCD-based TI system, known for its robustness and is the candidate for fault-tolerant quantum computation \cite{murali2019full, foss2024progress, kang2024seeking}, does not have much to offer. Prior work on shuttling-trapped ion prefers to evaluate the hardware architectures while using some compilation schemes \cite{murali2020architecting,schoenberger2024hardware, brown2016co}. There are also prior works that try to solve the circuit mapping problem for shuttling-based TI systems such as \cite{schoenberger2024shuttling, schoenberger2024using, tseng2024satisfiability, ovide2024scaling, durandau2023automated}. However, each has its downside. Tseng \textit{et al.} and Durandau \textit{et al.} \cite{tseng2024satisfiability, durandau2023automated} only considers minimizing shuttling operations in a one dimension system (linear segment). The works from Schoenberger \textit{et al.} \cite{schoenberger2024shuttling, schoenberger2024using} attempt to perform the circuit optimization for the operations, but the approach is hardware-agnostic. They simplify the hardware architecture into a memory zone and a processing zone. Their memory zone is a geometric grid composed of uniform blocks, and the main target is to move the ions from the memory zone to the processing zone while resolving congestion by cycle-based shuttling. This simplification can not capture the big picture yielded by the 2D QCCD architecture. For example, given many executable traps that we can use, the shuttling problem is not simply to move the ions to a specific processing zone but also to find the optimal trap zone for each ion. Moreover, this simplification fails to catch the congestion caused by having multiple trap zones.
 
 In \cite{ovide2024scaling},  Ovide \textit{et al.}  attempt to find the best initial mapping from the device's architecture, but do not solve the scheduling problem. Besides direct attempts to solve the circuit mapping problem as listed, there are efforts to solve the whole compilation framework. For example, Kreppel \textit{et al.} \cite{kreppel2023quantum} try to find the best-compiled quantum circuit to run with TI hardware but does not touch the shuttling operations, while Saki \textit{et al.}\cite{saki2022muzzle} give an optimized version of compiled circuit and shuttling sequences but only deals with one-dimensional architecture. Dai \textit{et al.}\cite{dai2024advanced} propose a probabilistic formula for ion movement and heuristic local layer generation and layer compression, which faithfully follows the QCCD architecture and yields a valid shuffling schedule. We acknowledge that Zhu \textit{et al.} \cite{zhu2025s} introduce a similar graph concept as our position graph and SABRE-inspired heuristic, but they do not try to resolve the congestion problem when the architecture is filled with a lot of ions.
 
In conclusion, to bridge these gaps, our work (1) enriches the works of QCCD-based TI compilation by bridging the difference between TI and superconducting-based literature with the \textit{position graph} abstraction, (2) finds the best-compiled circuit on the given two-dimension architecture with executable traps, loading zones, shuttling paths and junctions, (3) proves that the \textit{position graph} abstraction works well by coming up with SHAPER and SHAW to orchestrate the ions with shuttling operations, and finds the best-permuted circuit while resolving the congestion and deadlock problems.

%% file: isca2024/content/2background.tex
\section{Quantum Compilation}
\label{sec:QIC}

The fundamental information unit of quantum information is a quantum bit or \textit{qubit} for short, which represents a quantum state that belongs to the Hilbert space spanned by two basis states. For computation, these basis quantum states are $\ket{0}$ and $\ket{1}$, represented by the vectors $\left[ \begin{smallmatrix} 1 \\ 0 \end{smallmatrix} \right]$ and $\left[ \begin{smallmatrix} 0 \\ 1 \end{smallmatrix} \right]$, respectively. A qubit, $\ket{\psi}$ is then represented as a superposition of these: $\ket{\psi} = \alpha\ket{0} + \beta\ket{1} = \left[ \begin{smallmatrix} \alpha \\ \beta \end{smallmatrix} \right]$, where the amplitudes $\alpha, \beta \in \mathbb{C}$ and $\abs{\alpha}^2 + \abs{\beta}^2 = 1$. When measuring the quantum state, the probability of receiving state $\ket{0}$ is $\abs{\alpha}^2$, and state $\ket{1}$ is $\abs{\beta}^2$, respectively. Composing multiple qubits together in an $n$-qubit system creates a superposition state over all n-bit strings, formally defined as $\ket{\Psi} = \sum_{i=0}^{2^{n}-1} \alpha_{i} \ket{i}$ where $\alpha \in \mathbb{C}^{2^{n}}$ and $\sum_{i=0}^{2^{n}-1} \abs{\alpha_{i}}^{2} = 1$. After measurement, state $\ket{i}$ is received with probability $\abs{\alpha_{i}}^{2}$.

A quantum state is transformed by matrix operators that maintain the previously mentioned constraints. By definition, these linear transformations belong to the unitary group $U(2^n)$, where $n$ is the number of qubits transformed. While any \textit{unitary} is a valid state transformation, each quantum hardware architecture provides a small, fixed set of natively executable operations due to engineering and physical constraints. Quantum computers that are \textit{universal} can implement any operation with a sequence of their native instructions. These native instructions are typically referred to as \textit{gates} and described compactly by small unitary operators. In the context of the trapped-ion system, the native set varies slightly across vendors. 


Finally, quantum programs are expressed in the \textit{circuit model} consisting of (1) initialization of qubits depicted as wires going left to right through time, (2) quantum gates evolving the state of the qubits they touch, and (3) measurements resulting in the final readout. Without considering the measurement and initialization, the complete function a quantum circuit implements can be represented by the unitary matrix $U$ resulting from properly multiplying all gate operations~\cite{nielsen2001quantum}. We say that two quantum circuits are equivalent if they implement the same unitary matrix $U$ up to a \textit{global phase} -- a complex factor with magnitude one.

\subsection{Quantum Compilation}
In the context of quantum computing, the term ``compilation" often refers to the process of transforming a given ``logical'' quantum circuit (i.e., the quantum algorithm) into an equivalent, natively-executable circuit given a specific quantum processing unit (QPU). Compilers must overcome challenges such as limited qubit connectivity, gate fidelity, and other hardware-specific constraints in this essential task. Additionally, effective compilation enhances the performance of quantum algorithms and reduces errors by minimizing resource utilization. The Noisy Intermediate-Scale Quantum (NISQ) era requires these steps, as high gate error rates and limited coherence times currently characterize devices. Therefore, compilation's primary goal is to reduce instruction counts and execution time. The two main processes in quantum compilation are transpilation and circuit mapping.

\paragraph{Transpilation}
The goal of transpilation is to retarget the original circuit into one that only uses gates from the native gates set given a target QPU. In quantum computing, each platform has its own set of universal gates. For instance, superconducting platform \cite{koch2007charge} has the universal gate set of $SX$, $RZ$, and $CNOT$, while trapped-ion platform \cite{kielpinski2002architecture} has the universal gate set of $RZ$, $U1q$, and $RZZ$. State-of-the-art methods for transpilation involve re-synthesizing partitioned subcircuits from a larger circuit~\cite{younis2022quantum}. Here, large circuits are first partitioned into blocks of fixed sizes by grouping adjacent gates. Then, the unitaries for each block are calculated and re-synthesized in the new gate set using one of many different algorithms.

\paragraph{Circuit Mapping}
Circuit mapping, also known as routing and layout, is responsible for overcoming the connectivity constraints of a quantum architecture. This compilation step is usually mentioned in the context of superconducting architectures, 
where physical qubit connections are defined by a coupling graph, and multi-qubit gates can only be physically operated between connected qubits. Therefore, routing operations, such as a SWAP gate, must be chained to map a physically distant logical operation to a sequence of physical gates. Although necessary, this compilation procedure can greatly increase the execution time and gate count of a program and, therefore, is the most important to optimize.
To minimize the overhead, circuit mapping algorithms such as \cite{li2019tackling, bhattacharjee2019muqut, molavi2022qubit} are used to (1) derive the best initial mapping from \color{teal}program qubits \color{black} (from the original circuit) to the physical qubits (from the hardware architecture) and (2) perform intermediate mapping transition to remap the far-distance qubits to physically-coupled qubits.
In the QCCD-based TI platform, the shuttling process is similar to this mapping process. The difference is that shuttling utilizes a variety of ion shuttling operations, which physically move the ion, rather than logical SWAP gates, which mathematically move a qubit's state.

\subsection{Permutation-Aware Mapping}

The \emph{permutation-aware mapping} (PAM) is a state-of-the-art technique for mapping quantum circuits to superconducting devices~\cite{liu2023tackling}. It is based on the topology- and permutation-aware synthesis principles, as well as the mapping heuristic, such as SABRE \cite{li2019tackling}. We expand on these principles in this work and, therefore, introduce them here.

\label{sec:PAS_and_PAM}
\paragraph{Topology-aware synthesis} Quantum circuit synthesis finds a quantum circuit that implements a given target unitary matrix. This process is made topology-aware when the resulting circuit is already mapped to a target QPU's topology or coupling map. The topology-aware quantum circuit synthesis algorithms in recent years have the following intuition: by ``enumerating" the space of possible solutions, in our case, the space of possible circuits, the desired solution can be found by searching the space. Specifically, Qsearch \cite{davis2020towards} performs the space enumeration through tree construction. Each branch constitutes a possible gate operation on specific qubits with the root node as the initial layer. Branches not directly mappable to hardware are pruned, providing the topology-awareness. After the construction, a search algorithm is deployed to find the desired node in the tree that satisfies the optimality criteria. As Qsearch steps over a candidate solution, it employs a numerical optimizer to find the best parameters for that structure.

\paragraph{Permutation-aware synthesis} Following this line of research, \cite{liu2023tackling} introduces permutation-aware synthesis (PAS). PAS makes two observations toward better synthesis. First, classically permuting qubit indices during initialization and readout is a trivial process. Second, given a unitary matrix $U$, one of its classically-resolvable permutations, $P_{o}UP_{i}$ for some permutations $P_o, P_i$, may be easier to synthesize and require fewer quantum resources. To this effect, when synthesizing a unitary matrix $U$, it will enumerate all possible permutations and introduce them in the circuit. The permutations are introduced as an identity since $PP^T = I$ for all permutations $P$. This results in $P_{o}^{T}P_{o}UP_{i}P_{i}^{T} = U$. The outer permutations are factored out to the qubit initialization and readout stages, and $P_{o}UP_{i}$ is left to synthesize. This process produces very resource-efficient circuits and, in some cases, leads to circuits using fewer resources than provable-optimal circuits since it breaks the assumptions of the proof.


\paragraph{Heuristic mapping}
Heuristic mapping algorithms walk the circuit gate-by-gate and insert swaps determined by a heuristic to make the next gate or set of gates executable. The canonical algorithm is SABRE~\cite{li2019tackling}, which divides the circuit into the front and extended layers. The \emph{front layer} consists of the minimal nodes in a topological ordering, and the \emph{extended layer}, representing a lookahead window, consists of some configurable number of gates after the front layer. Its heuristic determines which swap to insert next by balancing the routing cost for the gates in the front layer with gates in the extended layer. To perform layout, this routing procedure is repeated in reverse, often multiple times, with the final ordering providing a good starting layout.


\paragraph{Permutation-aware mapping}

Permutation-aware mapping (PAM) \cite{liu2023tackling} combines the concepts of heuristic mappers with permutation-aware synthesis. To introduce PAS into the mapping process, a large circuit is first partitioned into small synthesizable blocks. These blocks are then synthesized for every permutation and stored. The SABRE mapping algorithm then proceeds with a generalized heuristic for blocks rather than gates. When a block in the front layer is made executable and moved off the front layer, another heuristic is performed that selects the best permutation of the block. PAM resolves the introduced outer permutations by merging them with the current state of the qubit mapping. As a result, PAM's heuristic balances the number of gates in the synthesized permutation with the effect the permutation will have on the rest of the mapping process.

%% file: isca2024/content/2QCCD.tex
\section{Quantum Charge-Coupled Device}
\label{sec:QCCD}
Trapped-ion (TI) platforms for quantum computation have drawn significant attention due to their robustness and seemingly all-to-all connectivity, leading to their widespread use in the NISQ era. Compared with its competitors, such as superconducting-qubit systems, the TI system is known not to suffer from the connectivity challenge, which is a massive hindrance in superconducting systems \cite{kjaergaard2020superconducting}. However, trapped-ion architectures face a scaling issue with the current linear chain architecture. \color{teal} Considering a linear trap, when the max trap capacity increases (number of maximum ions stored in the trap), \color{black} so does its heating when performing instructions. This presents difficulties in ion control and gate implementation for long ion chains and is directly related to a system's fidelity~\cite{wu2018noise}.

To tackle this problem, the Quantum Charge Coupled Device (QCCD) model was proposed~\cite{kielpinski2002architecture} as a modular and scalable trapped-ion architecture, where small traps are connected through ion shuttling paths. In more detail, the QCCD-based TI system consists of multiple small-capacity traps, as well as segments and junctions providing pathways for the ions to move from trap to trap. Figure~\ref{fig:H_QCCD} illustrates an example of a QCCD architecture. By restricting the trap size capacity, QCCD design achieves fast, high-fidelity two-qubit operations within each trap. Figure~\ref{fig:H_QCCD} illustrates the shuttling process as an ion physically moves from one trap to another. Currently, many state-of-the-art TI systems \cite{moses2023race, mordini2024multi, pino2021demonstration, sterk2024multi, delaney2024scalable} and the future development roadmap of TI vendors \cite{quantinuum2023roadmap} are built based on the concept of QCCD.


Effectively, the QCCD design trades the scaling issue for a connectivity challenge, as the all-to-all model is only available for ions within the same trap. This trade-off is desirable as the scaling issue prevents long-term adoption, whereas resolving shuttling schedules becomes a compiler stage. However, hardware vendors have continued to uphold an illusion of all-to-all connectivity between any qubit~\cite{quantinuum_systems}. This creates an inefficient compiler stack as, first, a program is optimized without concern for the QCCD architecture. Only after this stage, is a shuttling sequence resolved. This can lead to inefficient circuits since the first step may perform a seemingly great optimization, which will lead to more shuttling operations in the second compilation stage. Since shuttling operations account for a majority of the compute time, they must be minimized with the highest priority. For example, a movement operation between junctions may require around 100 to 120 microseconds \cite{gutierrez2019transversality}, whereas a gate operation only requires 30 to 100 microseconds \cite{gaebler2016high, ballance2016high}.

\begin{figure}
    \centering
    \includegraphics[width=0.6\linewidth]{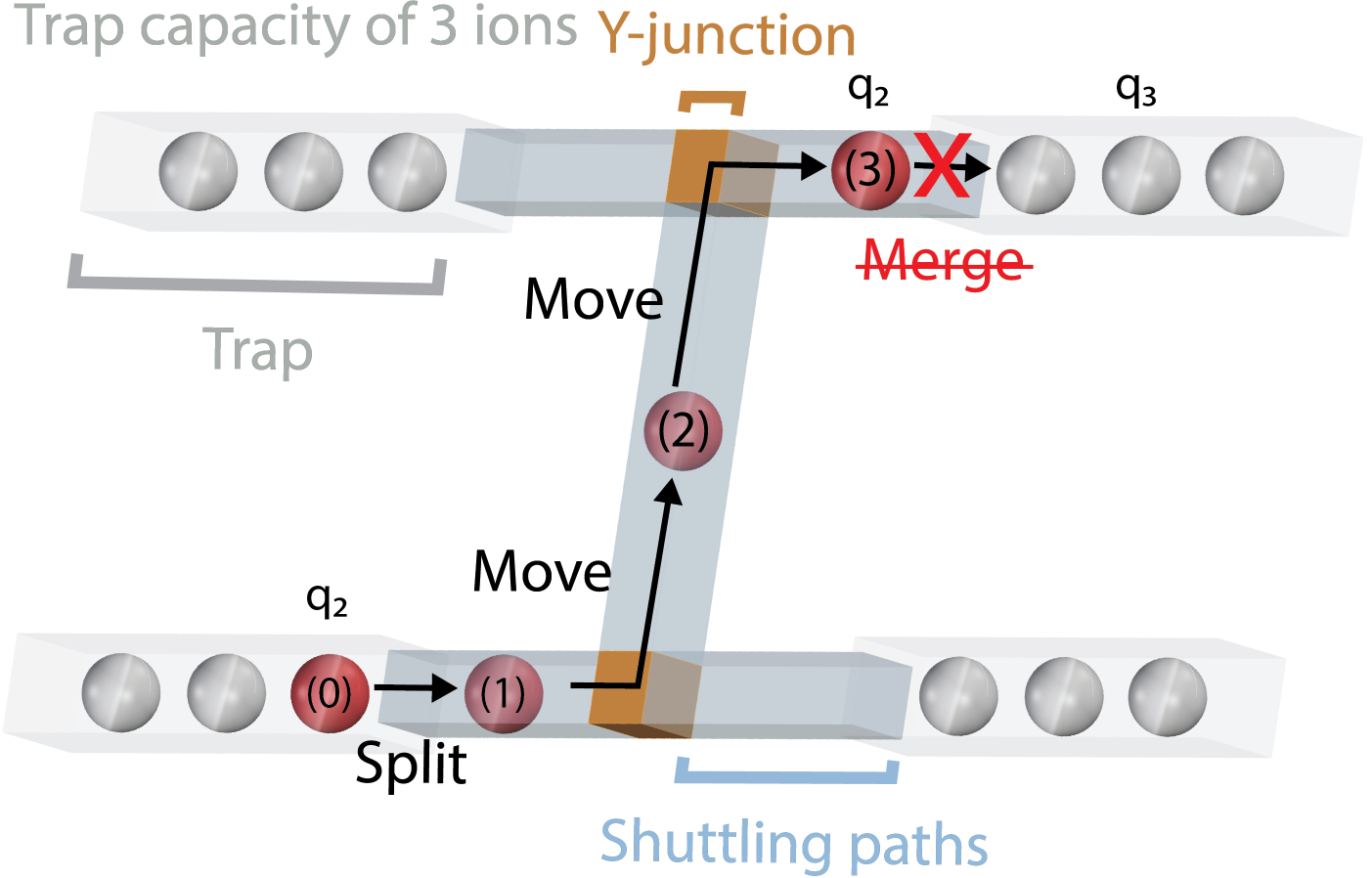}
    \caption{An example of congestion created when trying to move $q_2$ to the same trap with $q_3$. Here, after moving $q_2$ next to the trap containing $q_3$, the merge operation fails due to not enough free space in the trap. The congestion is created as $q_3$ can not be merged into the trap, and there is no one-step shuttling operation to free space on the $q_3$ trap.}
    \label{fig:congestion}
\end{figure}

Moreover, QCCD architecture also introduces specific constraints related to gate execution. As depicted in Figure \ref{fig:H_QCCD}, ions must be moved to an executable zone for gate operations. To represent this constraint, we assume two trap types exist: executable and storage traps. Therefore, the following constraints are established: \color{teal} \textbf{(1)} single-qubit gates can be applied to ions in executable traps, and \textbf{(2)} multi-qubit gates can be applied only to ions in the same executable trap. \color{black}
Besides these constraints, moving physical ions between traps using shuttling operations also implies restrictions \cite{murali2020architecting, schoenberger2024shuttling,malinowski2023wire} that we specify as follows 
\begin{enumerate}
    \item Only one ion can exist on a segment simultaneously.
    \item At a given time, only one ion can pass through a junction.
    \item Traps cannot contain more ions than their maximum capacity.
    \item The \emph{split} operation removes the outermost ion from a trap into a connected segment.
    \item The \emph{merge} operation adds an ion from a junction to the outermost position in a trap.
    \item The \emph{move} operation moves an ion from one segment to another, passing through a junction.
    \item The \emph{shuttling} operations can be executed in parallel as long as there is no collision.
    \item Gate operations on different traps can be executed in parallel.
\end{enumerate}
The above restrictions are usually not considered during the circuit optimization stages due to their complexity in resolving congestion. Figure~\ref{fig:congestion} depicts an example of congestion when trying to move $q_2$ to the same trap as $q_3$. In this case, we unintentionally create congestion as $q_2$ cannot be merged into the trap, and free space cannot be made trivially. 

%% file: isca2024/content/3position_graph.tex
\section{Position Graph abstraction}
\label{sec:position_graph}
We introduce a novel abstraction called the \emph{position graph}, which abstracts the physical architecture of superconducting and QCCD-based TI systems using the graph language. Numerous qubit mapping algorithms for superconducting devices can be trivially ported using this abstraction, enabling them to reason about the QCCD model's complex constraints directly. The position graph breaks the all-to-all connectivity assumption, offering a more expressive representation of the shuttling constraints in a structural and meaningful way. An example is illustrated in Figure~\ref{fig:position_physical}. Given the hardware architecture, the position graph is an abstraction that can fully express a device's architecture and naturally enforce its constraints. 
\\
Let us formally define it. 
\begin{definition}[Position Graph] 
\label{def:position_graph}
A Position Graph is a directed graph given by $G(\mathbf{V}, \mathbf{E}, \psi_v, \psi_e)$ where $\mathbf{V}$ is the set of vertices, $\mathbf{E}$ is the set of edges, $\psi_{\textbf{V}}: \mathbf{V} \rightarrow L_V$ is a labeling function that assigns labels to vertices, and $\psi_{\textbf{E}}: \mathbf{E} \rightarrow L_E$ is the labeling function that assigns pair of labels (movement label and execution label) to edges. $$L_\mathbf{V} = \{\text{None}, \text{Execute}, \text{Measure}, \text{Execute+Measure}\}$$ $$L_\mathbf{E} = \{\text{None}, \text{Move}, \text{Swap}, \text{Move+Swap}\}\times\{\text{None}, \text{Execute}\}$$
A node $p\in \mathbf{V}$ corresponds to a possible position of a \color{teal} program qubit \color{black}. The operations allowed on a \color{teal} program qubit \color{black} in that position are captured by the vertex labeling function $\psi_\textbf{V}$.

An edge $ij\in \mathbf{E}$ corresponds to a connection at position $i\in \mathbf{V}$ to $j\in \mathbf{V}$. The types of movement and gate executions allowed on the edge are captured by the edge labeling function $\psi_\textbf{E}$. Some edges only allow gate execution and no movement, while others allow movement with or without execution. Two types of movement are necessary to abstract away hardware specifics: \textit{move} and \textit{swap}. The move operator allows one qubit in a position to go to an empty connected position. The swap operator allows two qubits occupying connected positions to exchange their positions.
\end{definition}

\subsection{Encoding Superconducting Systems}
\label{sec:sc_encoding}
For superconducting systems, the position graph is a generalization of the coupling graph and potentially provides a unifying abstraction over different hardware architectures. From Definition~\ref{def:position_graph}, coupling graphs can be expressed using the position graph abstraction with the number of positions equal to the number of physical qubits, all positions being executable and measurable, and all edges being swappable and executable. With this abstraction, the state of the coupling graph, where \color{teal} program qubits \color{black} are moved around with \textbf{Swap}, can be fully captured with the position graph. Therefore, any superconducting system that can be represented with a coupling graph is also representable with a position graph, and any algorithm that can compile using the position graph abstraction can compile for superconducting systems.

\subsection{Encoding QCCD-based Trapped-Ion Systems}
\label{sec:ti_encoding}
\begin{figure}
    \centering
    \includegraphics[width=0.55\linewidth]{isca2024/figures/QCCD_ABSTRACTED.pdf}
    \includegraphics[width=0.44\linewidth]{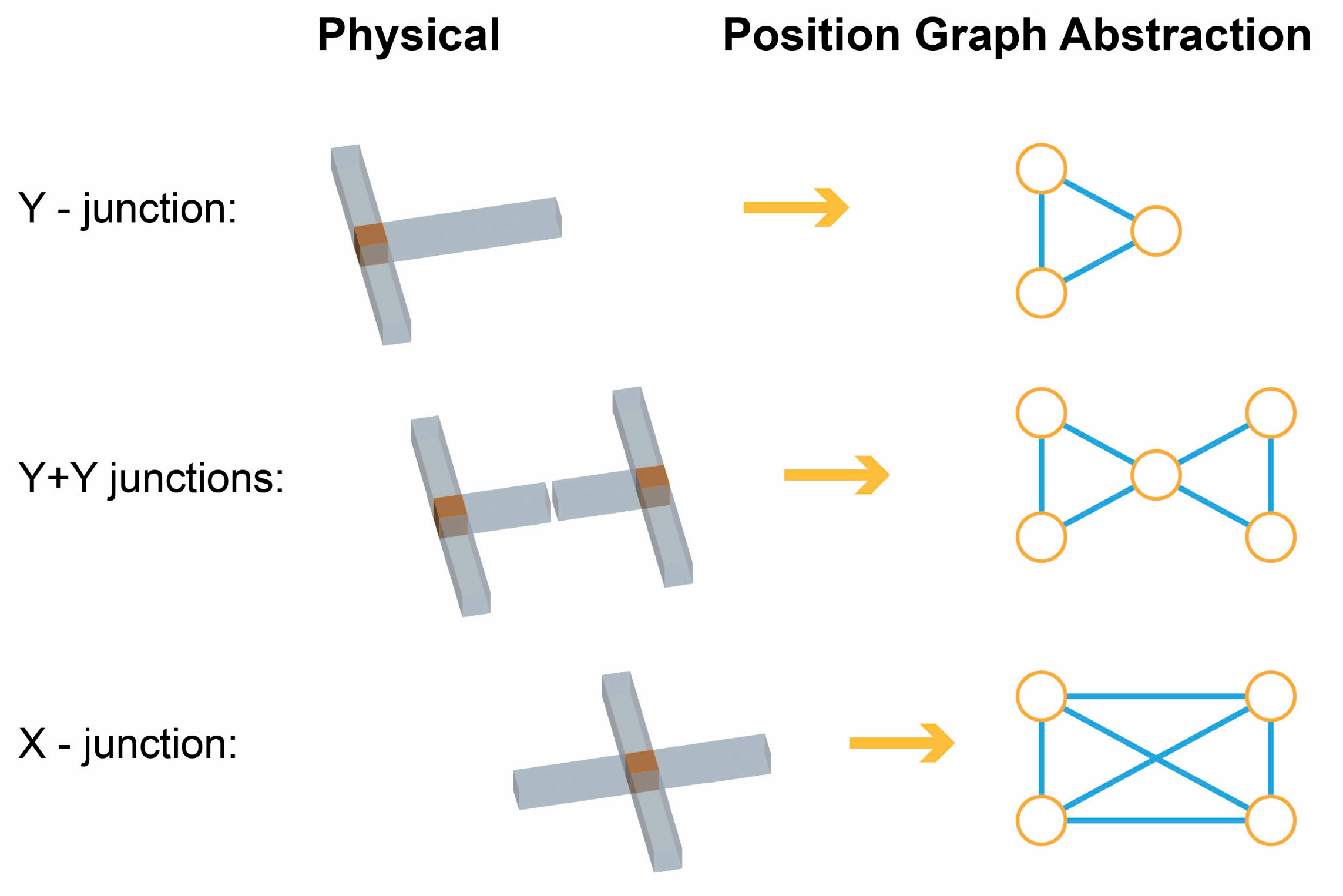}
    \caption{Example of mapping a QCCD-based TI architecture to its corresponding position graph.
    \color{teal}Figure (a) shows a real hardware configuration, and Figure (b) shows the position graph abstraction that represents the hardware configuration. Here, (i) each \textit{position} vertex in the position graph represents a physical location where an ion may reside. The vertex colored purple indicates that an ion is currently occupying this location, while the uncolored (white) vertex indicates that this position is empty. (ii) Each vertex with a black border represents the available position inside a trap, while the vertex with a yellow border represents abstract intermediate positions where a qubit may temporarily reside while moving between traps. (iii) The dotted black edges within each trap indicate pairs of positions between which a two-qubit gate operation is possible, while the straight line indicates the physical connection that allows operations such as swap. (iv) The blue edges abstract the path where the ion can move (without stopping) to reach another trap.
    \color{black} Additionally, the right figure shows how the Y-junction (T-junction), X-junction, and combination of Y-junction (T-junction) are transformed into abstraction.}
    \label{fig:position_physical}
\end{figure}
For QCCD-based systems, the position graph allows compilers to break the all-to-all connectivity assumption, offering a more expressive representation of the shuttling constraints in a structural and meaningful way, as shown in Figure~\ref{fig:position_physical}. In more details, for the QCCD-based TI system, Definition \ref{def:position_graph} treats each node of the position graph not as an ion or qubit like the physical connectivity graph, but rather as a possible position that an ion or qubit can occupy. Similarly, shuttling paths are captured by nodes or positions with constraints on their movement through edge labels. This naturally enforces all constraints mentioned in Section~\ref{sec:QCCD} and aims to avoid congestion. 

From a physical QCCD architecture, the position graph can be easily constructed by consecutively going through all traps, junctions, and shuttling paths. There are three stages to encoding traps into a position graph. The first step is to connect neighboring ions in a trap with a movable and swappable edge because only adjacent ions can be physically swapped. Second, all ions are connected to all others in the same trap with an executable edge. This is because logical two-qubit operations can be applied to any pair of ions in the same trap. Lastly, all individual positions are marked as executable and measurable. These steps apply only to executable traps; storage-type traps only require the first step.

Next, we iterate through the shuttling paths. All edges in the shuttling path are only given the \textbf{Move} capability, as gates cannot be applied to an ion in transit and are not swappable outside of a trap. We consider each segment a position since it represents a point where ions can exist throughout the shuttling paths. A $d-$degree junction creates a complete subgraph with $d$ nodes for the position graph. The motivation behind this is that a $d-$degree junction creates all-to-all connections between $d$ positions. That is, it allows pairwise connectivity between all connected segments. After this, we need to create edges that connect the disconnected components (traps and traps, traps and junctions) with respect to their connections in the physical architecture (the segments). The edges that connect the components are assured to reflect the shuttling operations.

Given the physical QCCD architecture with (1) the set of ion traps $\mathbf{T}$, $|\mathbf{T}| = \mathbf{m}$, where each trap $t\in \mathbf{T}$ has a capacity of $\mathbf{n}_t$ 
ions, (2) the set of segments $\mathbf{S}$, and (3) the set of junctions $\mathbf{J}$. 
Therefore, 
\begin{align}
    \abs{\mathbf{V}} &= \sum_{t\in \mathbf{T}}\mathbf{n}_{t} + \abs{\mathbf{S}}\\
    \abs{\mathbf{E}} &= \sum_{t\in \mathbf{T}}\mathbf{n}_{t}(\mathbf{n}_{t}- 1) + \sum_{\mathbf{j} \in \mathbf{J}} d(\mathbf{j})(d(\mathbf{j}) - 1) + 2\abs{\mathbf{JT}} + 2\abs{\mathbf{TT}}    
\end{align}
where $\mathbf{j}$ denotes a junction, $\mathbf{JT}$ denotes the set of segments connecting junctions and traps, $\mathbf{TT}$ denotes the set of segments connecting pairs of traps, and $d(\mathbf{j})$ denotes the degree of junction $\mathbf{j}$. Note that the edges are not bidirectional in the position graph abstraction; however, they are in QCCD, so every edge gets added twice, once for each direction. For simplicity, the figures in this paper show bidirectional edges without arrows. As an example in Figures \ref{fig:position_physical} and \ref{fig:position_physical}, QCCD architecture contains 4 traps, each with a maximum capacity of 3 ions, the set of junctions $J$ contains 2 junctions, and there are 5 segments in ${\bf S}$. Thus, in this case, the hardware architecture can be abstracted to a position graph $G$ with 17 nodes and 44 edges (22 bidirectional).

\subsection{States of Position Graph}
\label{sec:state_of_position_graph}
\begin{figure*}
    \centering
    \includegraphics[width=\linewidth]{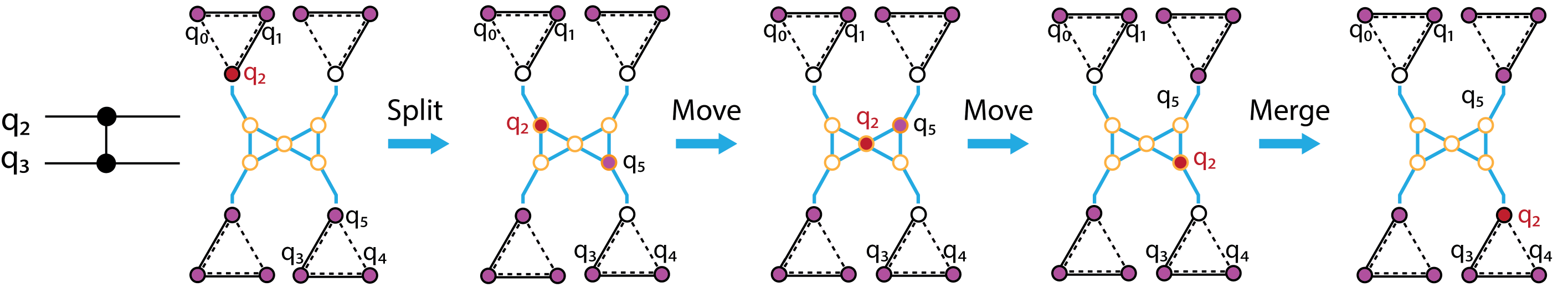}
    \caption{States of position graph during a sequence of shuttling operations. \color{teal} The ion representing \color{teal} program qubit \color{black} $q_{2}$ is color as red. As we want to execute a two-qubit gate acting on $q_{2}$ and $q_{3}$, the two ions representing those need to be moved to the same trap. The transitions are shown as follows: (1) A split operation to split $q_2$ and $q_5$ out of their current trap, (2) a sequence of moves to avoid collision and congestion, (3) a final merge to move $q_2$ to the trap containing $q_3$, where the gate can be applied.
    \color{black}.
    The two-qubit gate acting on $q_{2}$ and $q_{3}$ requires several transitions, each reflecting a state change. Whenever an operation moves $q_{2}$, the position graph fully captures the new state.} 
    \label{fig:state_of_position}
\end{figure*}

Given a position graph $G_p$, the assignment of \color{teal} program qubits \color{black} to physical positions, $\Phi$, constitutes the state of the position graph.
The qubit assignment $\Phi: \mathbf{Q} \rightarrow \mathbf{V}$, where $\mathbf{Q}$ is the set of \color{teal} program qubits \color{black}, maps all \color{teal} program qubits \color{black} from the original quantum circuit to a position in the graph. Legal shuttling operations change the states. For example, Figure~\ref{fig:state_of_position} displays the change of state in the position graph $\Phi$ caused by the change in ion assignment after each shuttling operation when moving $q_{2}$ to the trap containing $q_{3}$.. 

Specifically for the QCCD-based TI system, the position graph enforces that no ion swap can happen outside of the trap zone. This naturally forms the constraint of junctions and shuttling paths, and reflects congestion. An example of how the position graph exhibits congestion and imposed constraints is also shown in Figure \ref{fig:state_of_position}. When moving $q_2$ to the trap that contains $q_3$ for gate execution,  a blockage with $q_5$ will surely create congestion if not moved, as demonstrated in Figure \ref{fig:congestion}. This is clearly exhibited by the architectural abstraction from the position graph where there is no valid move to trap containing $q_3$.

\subsection{Routing for QCCD-based architecture using Position Graph}
\label{sec:theoretical_foundation_pg}

When considering the ion routing problem using position graph abstraction, we realize its theoretical connection to the well-known problems called ``Token Reconfiguration problem'' (TRP), also known as a version of the Pebble Motion problem, \cite{cualinescu2008reconfigurations,kornhauser1984coordinating} in theoretical computer science and network science. For completeness, we define the problem as follows: Let $G(\mathbf{V}, \mathbf{E})$ be a graph with $\abs{\mathbf{V}} = n$ vertices with set of tokens $\mathbf{P} = \{1, \dots, k\}$, where $k < n$, on distinct vertices. A configuration of tokens on $G$ is a mapping $S: \mathbf{P} \rightarrow \mathbf{V}$ such that $S(i) \neq S(j)$ if $i \neq j$. A transfer of token $p$ from vertex $u$ to adjacent vertex $v$ is denoted as $(p, u, v)$ and is only valid if and only if vertex $v$ is unoccupied. The Token Reconfiguration problem is to find the shortest sequence of transfers to reach out the final configuration $S_{+}$ from the initial configuration $S_{0}$ or determine that $S_{+}$ is unreachable. 

This problem is well-studied in the theoretical computer science community and is known to be NP-hard \cite{ratner1990n2} and APX-hard \cite{cualinescu2008reconfigurations}. By treating ions as tokens or pebbles, we can directly map the notation of the position graph into with some subtle adaption. Specifically, the position graph state (ion assignment) $\Phi$ can be considered as configuration $S$, \textbf{\textit{merge/split}} and \textbf{\textit{move}} labeled edges can be treated as normal edges, while \textbf{\textit{swap}} labeled edges can be treated as a variant where tokens/pebbles are interchangeable \cite{papadimitriou1994motion}. \color{teal} Here, we consider the simplest version of our problem, where we want to execute a single gate. Let consider the example of executing a two-qubit gate ($q_{2}$, $q_{3}$) in Figure \ref{fig:state_of_position}. In this example, our initial configuration $S_{0}$ is the first position graph state, while our final configuration $S_{+}$ is the final position graph state where ions representing $q_{2}$ and $q_{3}$ share the same trap. Hence, the TRP problem fully captures the sequence of shuttling operations.
\color{black}
Although the simplest version of our problem can be directly cast as the Token Reconfiguration, we emphasize that standard, well-studied token-based approaches do not trivially solve our problem because our target is to shuttle the ions into the global trap space \emph{where the final configuration is unknown}.

\color{teal}
The shuttling problem in full detail requires the ions to be shuttled so that all gates are executed while minimizing runtime. Therefore, at any given time step, the final ion configurations $S_{+}$ are decided due to the choice of which gates to execute and which traps to select that potentially minimize the runtime (this is similar to the objective of job scheduling, where we choose which job to send to which machine). Therefore, it can be thought of as a generalization with an inner loop using TRP to perform the shuttling to move the ions to the final configuration, whereas the outer loop decides the best final configuration for the inner loop based on the gate and architecture (job scheduling). 
\color{black}
For completeness, \emph{we derive the mixed-integer linear programming (MILP)} of the shuttling problem of the TI-QCCD architecture as in appendix \ref{sec:shuttling_MILP}. 

In a relaxed manner where $k \ll n$, our problem can also be framed as cooperative pathfinding (multi-agent pathfinding, MAPF) \cite{silver2005cooperative, standley2010finding, luna2011push}, which has also been well-studied. Besides configuration selection, deadlocks can happen during shuttling when the amount of space in executable zones is less than the number of ions to execute \cite{dally1987deadlock} or bottleneck the shuttling path \cite{luo2003real,silver2005cooperative}. To ensure deadlock-free motion in future work, one could consider deadlock-free routing algorithms in network-on-chip (NoC) systems and computer networks \cite{ebrahimi2017ebda, deorio2012reliable, anjan1995efficient, domke2011deadlock, ogras2006s, van2010graph} and deadlock avoidance in MAPF \cite{chan2022multi, wang2020walk,sharon2015conflict} by customizing and adapting those to the position graph. For example, based on Dally's theory \cite{dally1987deadlock}, we can try to break the cyclic dependency between traps (when ions compete for traps) to resolve the deadlock. This bridging is non-trivial as the constraints yielded by the shuttling problem are non-standard from the perspective of MAPF and NoCs, and it is interesting to try adapting the well-known protocols to the shuttling problem through the abstraction of the position graph. We note that in this study, we use a heuristic method to resolve these congestion and bottlenecks and leave the investigation of using more advanced methods for future work.

%% file: isca2024/content/4QCCD_mapping.tex
\section{Position Graph Compilation}
\label{sec:our_algo}
We devise two novel mapping algorithms inspired by the well-known SABRE and PAM methods to provide example mapping algorithms for the position graph abstraction. Adapting these methods to handle the more general position graph, which enables them to resolve shuttling schedules for TI systems, requires primarily generalizing the mapping heuristic in both.



\subsection{Position Graph Heuristic Mapping}
\label{sec:pam_map}

Following the strategies of PAM, a partitioner first groups gates in the circuit into $k$-qubit blocks. Then, each block is resynthesized for each permutation for all possible topologies, as described in Section~\ref{sec:PAS_and_PAM}. Note for trapped-ion systems, the only topology possible is an all-to-all topology because each k-qubit block must execute entirely in the same executable space (trap zone), which is always all-to-all. \color{teal} Here, the objective for choosing the best synthesized block comprised of both gate counts and potential for faster shuttling moves as shown in Figure \ref{fig:concrete_ex}. Here, the term potential of better shuttling moves is calculated by how congested the shortest path is when trying to move an ion from trap $A$ to trap $B$. 
As the main effects of shuttling are heating, which causes worse gate fidelity as shown in the fidelity model in Section \ref{sec:exp_setup}, the synthesis objective is then balanced between minimizing gate counts and congestion ($\mathrm{gate\_count} - \mathrm{congestion}$). \color{black} 


For the shuttling, a qubit assignment $\Phi$ tracks the state of the mapping heuristic algorithm by mapping program qubits to positions in the graph. A distance matrix $D$ is calculated using the Floyd-Warshall algorithm. This matrix, $D$, provides the cost for each shuttling path in the graph. The partitioned circuit is then divided into a front layer $F$ and an extended layer $E$. The last step of the preparation is to create an empty mapped circuit.

\begin{algorithm}
    \DontPrintSemicolon
    \caption{SHAW/SHAPER}
    \label{alg:SHAPER}
    \KwInput{Front layer $F$, Circuit $C$, state of position graph $\Phi$, distance matrix $D$, position graph $G_{p}$}
    \KwOutput{Instruction list $\mathcal{I}$, final state of position graph $\Phi'$}
    $\mathcal{I}$  $\gets [\ ]$\;
    \While{$F \neq \emptyset$}
    {
        \textit{executable\_list} $\leftarrow$ all executable blocks from $F$\;
        \If{$\textit{executable\_list} \neq \emptyset$}
        {
            \For{\rm circuit block $B$ \textbf{in} \textit{executable\_list}}
            {
                
                Find the best permutation $\pi$ for block $B$ \tcp*{Used only in SHAPER} 
                Apply permutation $\pi$ to $B$ \tcp*{Used only in SHAPER} 
                Update state of position graph $\Phi$ with $\pi$ \tcp*{Used only in SHAPER} 
                Update $F$ with successors of executed blocks\;
            }
        }
        \Else
        {
            \textit{move\_scores} $\gets []$\;
            Compute $E \gets extended\_set(F, C)$\;
            Compute \textit{possible\_moves} $\gets obtain\_moves(\Phi, G_{p})$\;
            \For{\rm \textit{move} \textbf{in} \textit{possible\_moves}}
            {
                Update assignment $\Phi$ with \textit{move} as $\Phi_{move}$\;
                \textit{move\_scores}[\textit{move}] $\gets \mathbf{H}(F, E, \Phi_{move}, D, G_{p})$ \tcp*{Details in equation \ref{eq:H}} 
            }
            Select $\textit{best\_move} \gets \arg\min{(\textit{move\_scores})}$\;
            \If{\rm No \textit{best\_move} \textbf{or} detected repeated path}
            {
                Escape local minima with Algorithm \ref{alg:local-min-res}\;
            }
            \Else
            {
                Append \textit{move} to $\mathcal{I}$  and Update $\Phi$\;
            }
        }
    }
\end{algorithm}


Unlike connectivity graph mapping used in SABRE for superconducting systems, we make use of the state of position graph $\Phi$, which reflects the mapping of the position of \color{teal} program qubits \color{black} (ion for QCCD-based TI case) as detailed in \ref{sec:state_of_position_graph}. The position graph state, reflecting ion assignment, shows the current state of all ions within the system and demonstrates the hard constraints of the system while combining them with the position graph. Here, the distance matrix $D$ contains all costs to shuttle between pairs of positions in the position graph,  calculated using Floyd-Warshall's algorithm \cite{cormen2009introduction}. \color{teal} In addition, the Floyd-Warshall's algorithm, which is good for dense small graphs, other well-known algorithms such as Dijkstra, Johnson algorithm, and their parallel implementation can be employed for large graphs \cite{pradhan2013finding}.\color{black}

Both algorithms proceed with their heuristic sweep similarly to SABRE and PAM. Algorithm~\ref{alg:SHAPER} lists the pseudocode steps for SHAW/SHAPER's main loop. The algorithms continue until the front layer $F$ is empty. The first step of the main loop is to gather any executable blocks in $F$ and move them to the mapped circuit. A block in $F$ is executable if all of its qubits are currently mapped to positions that lie on a connected component composed entirely of executable edges. For TI systems, this would imply that all of the block's qubits are currently in the same trap. SHAPER will additionally select a permutation, $\pi$, and associated synthesized circuit for the block, which updates the qubit mapping and becomes the block in the mapped circuit. SHAPER selects permutations using the same heuristic as in PAM~\cite{liu2023tackling}.

If there are no executable blocks in $F$ currently, shuttling instructions must be inserted to transition the position graph to a state with executable blocks. Given the current state, the next transition is selected by assigning a heuristic score to candidate states. While a heuristic is used in SABRE and PAM to choose the placement of the next swap instruction, this heuristic must be significantly adapted and is the main novelty in these algorithms. Equation~\ref{eq:H} provides the heuristic as two terms: $F$ for the contribution from the front layer and $E$ for the contribution from the extended set. The extended set provides look-ahead capabilities to the mapping algorithms and can have its effect adjusted by the weight $W_E$. In both formulations, $F$ and $E$, the first term in the summation indicates the maximum distance between all pairs of qubits from block B. Here, we use the maximum pairwise distance, $\max_{q_i, q_j \in B}  D$,  to ensure that the furthest qubit is always moving first. This property is valuable in position graphs, especially in TI systems. For example, if considering the sum of all pairwise distances, $\sum_{q_i,q_j \in B} D$, and most qubits are near each other with one faraway, the optimization landscape may become plateaus in most directions. For the second term, we take the sum of all qubit positions to their nearest executable position to ensure movement to executable positions.

We use $d(p) = \min_{T_{i} \in \mathbf{T}}d_{T_{i}}(p)$ to denote the distance from position $p$ to the nearest executable zone (trap). The distance from position $p$ to zone $T_i$, $d_{T_i}(p)$, is defined as follows: if zone $T_i$ has free space then $d_{T_i}(p) = \min_{space \in T_i}{(D[p][space])}$; otherwise, $d_{T_i}(p) = \infty$. This formulation does not just inform how the next move contributes to gate execution but also implies the state of the system. For example, in the QCCD-based TI case, when all traps are crowded, and some ions need to move to another trap, the split operation to move an ion from the trap gives the best cost. 
\begin{equation}
    \label{eq:H}
    \begin{aligned}
        \mathcal{F} &= \frac{1}{\abs{F}} \sum_{B \in F} \Bigg[ \max_{q_i, q_j \in B}  D[\Phi(q_{i})][\Phi(q_{j})] + \sum_{q_{i} \in B}d(\Phi(q_{i})) \Bigg]\\
        \mathcal{E} &= \frac{W_{E}}{\abs{E}} \sum_{B \in E} \Bigg[ \max_{q_i, q_j \in B}  D[\Phi(q_{i})][\Phi(q_{j})] + \sum_{q_{i} \in B}d(\Phi(q_{i})) \Bigg]\\
        \mathbf{H} &= \mathcal{F} + \mathcal{E},
    \end{aligned}
\end{equation}
where $W_{E}$ is the adjustable weight for the extended set and $\Phi(q_{i})$ denotes position  of qubit $q_i$ in block $B$ on $G_{p}$.

The scoring formulation from Equation \ref{eq:H} contains many local minima that can easily lead to a hold in our algorithm. Moreover, the SABRE algorithm is known to produce local minima and ways to escape from them. Therefore, we introduce a local-min resolution in \ref{sec:local-min} to help us escape from those local minima and avoid unnecessary moves. Moreover, as in the QCCD architecture, ions are not allowed to interact in the same shuttling paths; a mechanism to avoid being in a deadlock position caused by a bottleneck shuttling path is also introduced by pushing every ion back to the nearest traps. Heuristically, the order of gate execution is updated to allow the simplest gate to be executed first.  After performing shuttling operations to enable gate execution, we continue the loop until the front layer is empty. This indicates we have already traveled one pass through the circuit. By doing this pass again but in a reverse manner (perform SHAPER but with the circuit flipped), we end up with an initial mapping of ion assignment for the original circuit. This is usually called the layout process~\cite{murali2020architecting}.

\subsection{Escaping local minimum}
\label{sec:local-min}
As the optimization landscape contains many local minima, we may need to help the mapping algorithm escape the local minima while orchestrating the qubits. This escape process is described in Algorithm~\ref{alg:local-min-res}. Here, the block $B$ is given as input, and the physical positions of each \color{teal} program qubits \color{black} from block $B$ are retrieved through the state of the position graph, $\Phi$. First, the target executable component, $T$, is selected with respect to the distance (shortest path) and the congestion of the path capturing how hard it is to move the qubits to the space. Then, the qubit order $\chi$ is decided based on the same criteria to select $T$, where the first qubit to be moved to $T$ is the one with the minimum cost (shuttling time).


After selecting the executable zone and qubit order, the algorithm starts with finding the shortest path from $\Phi{(q_{i})}$ to the allocated space $T$ called $P_{\Phi(q_{i}) \rightarrow T}$ (this can be done through Dijkstra's algorithm \cite{dijkstra2022note}). Algorithm~\ref{alg:congestion_res} is called if (1) the selected space $T$ is filled, or (2) a qubit gets stuck in congestion during movement. If (1) then some space is released by moving the qubits at the endpoint of $T$, denoted as $EP(T)$. If (2), the congestion is found and cleared.
\begin{algorithm}
    \DontPrintSemicolon
    \caption{Escape Local Minimum}
    \label{alg:local-min-res}
    \KwInput{Block $B$, state of position graph $\Phi$, position graph $G_{p}$}
    \KwOutput{Instruction list $\mathcal{I}$, final state of position graph $\Phi'$}
    $\mathcal{I}$ $\gets [\ ]$\;
    \textit{physical\_positions} $\gets \Phi(q_{i})$ for $q_i \in B$\;
    Find best zone $T$ to move ions from \textit{physical\_positions} \tcp*{Based on distance from $\Phi(q_{i})$ to $T$} 
    Find the qubit order $\chi$ to move \tcp*{Move qubit with closest position first}
    \For{\rm qubit $q_{i}$ \textbf{in} $\chi$}
        {
        Find the shortest path $P_{\Phi{(q_{i})} \rightarrow T}$\;
        \If{\rm there exist no space in $T$}
            {
            $ resolve\_conges(EP(T), N(EP(T)), P_{\Phi{(q_{i})} \rightarrow T}, \Phi)$ \tcp*{Details in algorithm \ref{alg:congestion_res}}
            }
        \tcp{Move along the shortest path}
        \For{\rm move $P_{i \rightarrow i'}$ \textbf{in} $P_{\Phi{(q_{i})} \rightarrow T}$} 
        {
            \If{\rm there exits congestion in $P_{i \rightarrow i'}$}
                {
                $ resolve\_conges(i, i', P_{\Phi{(q_{i})} \rightarrow T}, \Phi)$ \tcp*{Details in algorithm \ref{alg:congestion_res}}
                }
            \Else
                {
                Perform move $(i \rightarrow i')$ and append to $\mathcal{I}$\;
                }
         }
        }
\end{algorithm}

In Algorithm \ref{alg:congestion_res}, the recursive resolution congestion algorithm tries to tackle any conflict when trying to move $p_{1}$ to $p_{2}$. As $p_{1}$ cannot move to $p_{2}$; $p_{2}$ must be occupied by an ion and both $p_{1}$ and $p_{2}$ must not be inside a swappable zone, or else they can perform swaps. To resolve this, $p_{2}$'s neighbors are considered to see if they are unoccupied and not lying on the main path $P_{\Phi{(q_{i})} \rightarrow T}$. If any neighbor of $p_{2}$ satisfies the condition, ion at $p_{2}$ is moved to that space, and ion at $p_{1}$ is moved to $p_{2}$ which resolved the congestion. Otherwise, we perform a recursive call with $p_{2}$ and its neighbor $N(p_{2})$ beside $p_{1}$, by doing this, we try to move the far-distance neighbor of $p_{2}$ away to create space for lower-distance neighbor and finally create space for $p_{2}$ to move out of the path. Note that this strategy can lead to a deadlock situation as a result of the presence of multiple ions outside of the trap zone. Therefore, we also enforce a mechanism to push every ion back to the nearest traps when the occupation rate on the shuttling path is too high.
\begin{algorithm}
    \DontPrintSemicolon
    \caption{Resolve congestion}
    \label{alg:congestion_res}
    \KwInput{Positions $p_{1}$ and $p_{2}$,
    path $P_{\Phi{(I)} \rightarrow T}$, state of position graph $\Phi$.}
    \KwOutput{Instruction List $\mathcal{I}$, final state of position graph $\Phi'$.}
    \tcp{Try moving qubit from $p_1$ to $p_2$}
    $\mathcal{I}$ $\gets [\ ]$\;
    Obtain neighbors of $p_{2} \gets N(p_{2})$ and remove $p_{1}$.\;
    \textit{moved\_flag} $\gets$ \textbf{FALSE}\;
    \For{\rm $p$ \textbf{in} $N(p_{2})$}
        {
        \tcp{If any neighbor of $p_2$ is empty and not involve in the path $P_{\Phi{(I)} \rightarrow T}$}
        \If{\rm $\Phi(p) == \emptyset$ \textbf{and} $p \notin P_{\Phi{(I)} \rightarrow T}$}
            {
            Perform move $(p_{2}\rightarrow p)$ and append to $\mathcal{I}$\;
            \textit{moved\_flag} $\gets$ \textbf{TRUE}\;
            \textbf{BREAK}\;
            }
        
        }
    \If{\textit{moved\_flag}}
        {
        Perform move ($p_{1}\rightarrow p_{2}$) and append to $\mathcal{I}$\;
        }
    \Else
        {
        \tcp{Start the recursive loop to move neighbor of $p_2$}
        $resolve\_conges(p_{2}, N(p_{2})\setminus p_1, P_{\Phi{(I)} \rightarrow T})$\; 
        }
\end{algorithm}
For better illustration, we show an example in Figure \ref{fig:state_of_position} of how our algorithm to resolve the congestions works in QCCD architecture. In this case, we want to move $q_{2}$ to the same trap zone of $q_{3}$, however, the trap containing $q_{3}$ is already full. Therefore, the first step is to move the ion at the trap's endpoint $q_{5}$ out of the trap, then we start moving $q_{2}$ toward the designated trap. When going to reach the trap, we can see that there is congestion caused by $q_{5}$, this can be easily solved by moving $q_{5}$ out of the path and toward another trap. Finally $q_{2}$ successfully merges into the same trap with $q_{3}$ and gate acting on $(q_2, q_3)$ is exectuted. 

%% file: isca2024/content/5evaluation.tex
\section{Evaluation}
\label{sec:evaluation}
\subsection{Experimental setup}
\label{sec:exp_setup}
The experiments are performed on well-known quantum algorithms and circuits ranging from 8 to 128 qubits. 
The Quantum Approximate Optimization Algorithm (QAOA)~\cite{farhi2014quantum} is a variational algorithm broadly applicable to NISQ devices for combinatorial optimization problems. The Quantum Fourier Transform (QFT)~\cite{nielsen2001quantum} is a common quantum circuit for designing larger algorithms such as the Quantum Phase Estimation~\cite{abrams1999quantum}. The Quantum Volume (QV) circuits are commonly used as a benchmark for evaluating quantum hardware performance~\cite{cross2019validating}. The Transverse Field Ising Model (TFIM) and Transverse Field XY (TFXY) circuits are quantum circuits for Hamiltonian simulation~\cite{shin2018phonon}. These are crucial simulation models for many near-term applications using quantum computing. Here, the QFT, QV, and QAOA are generated using the Qiskit framework~\cite{qiskit2024}. In particular, QAOA is generated based on Erd\H{o}s--R\'enyi graph with edge probability equal to $0.3$. The TFIM and TFXY circuits were generated by the F3C++ compiler~\cite{kokcu2022algebraic}. \blue{All experiments are performed on a 32-core AMD EPYC 7702 processor with 1TB of memory.}

We developed SHAPER and SHAW in Python~3.11 as a BQSKit~\cite{younis2021berkeley} extension. We evaluated both SHAPER and SHAW  against the state-of-the-art shuttling algorithm, QCCDSim~\cite{murali2020architecting}. Furthermore, we try the two-stage approach and perform a global optimization round using PAM/PAS \cite{liu2023tackling} with all-to-all assumptions before applying QCCDSim. To make the comparison more extensive, we made another contribution by formulating the shuttling problem of the QCCD-based TI architecture using MILP optimization model as shown in  Appendix \ref{sec:shuttling_MILP}. The MILP model is solved using the CP-SAT solver from OR-Tools \cite{cpsatlp} with the \textbf{timeout of $40,000$ seconds}. The result from MILP is then used to compare with other methods.\\

\color{teal}
QCCDSim only performs scheduling on two-qubit gates without any support for circuit synthesis. Our proposed method (SHAPER) performs both synthesis and scheduling. \emph{For fair comparison, we first transpile all circuits into Quantinuum's native gateset}, containing parameterized ZZ gates, $e^{-i (\theta/2) Z \otimes Z}$, and single-qubit Quantinuum U1q gate, through the BQSKIT compiler with optimization level $3$. We note that we do not add the compilation time to the QCCDSim benchmark, and this compilation scheme is pretty exhaustive, where further synthesis using the same optimization level compiler tends not to be very effective. Furthermore, we want to emphasize that this transpilation pre-processing is important, as without this, our circuit synthesis can easily reduce the number of gate counts by at least $2-3$ times, which makes the comparison between operation times trivial (we always outperform QCCDSim).
\color{black}
Our source code and results are available at [link will be added here upon acceptance].

As the formulation of MILP does not allow flexible timing (operation-specific durations) and the complexity of the program scales exponentially when we increase the circuit size, we only test MILP on small circuits. Accordingly, we adopt two timing models (see Table \ref{tab:timming}), for large circuits we keep native, operation-dependent hardware timings. \color{teal} This timing is obtained from \cite{gutierrez2019transversality}, where the two-qubit execution time is formulated as FM gates \cite{leung2018robust}. The details of timing for other shuttling operations are obtained
from real characterization experiments \cite{gutierrez2019transversality}\color{black}.
For small circuits, we switch to a discretized model with each tick (timestep) of $40 \mu s$. This discretization involves adding dummy nodes to edges to capture the total traversal time through given edge. As each time step is $40 \mu s$, all operation durations are rounded to a multiple of $40$. Noted that, the timing for the model can be characterized exactly with respect to the hardware, but it requires scaling up the model (hard or even impossible to solve). As a proof of concept, the chosen timestep is crucial for the CP-SAT solver to solve the MILP formulation. For fairness, when comparing with MILP, all heuristic use the same set of configurations.
\begin{table}
\begin{tabular}{|l|l|l|}
\hline
\textbf{Instructions} & \textbf{Small-scale circuit timing} ($\mu s$) & \textbf{Large-scale circuit timing} ($\mu s$) \\ \hline
Single-qubit gate execution    & 40                 & 30                       \\ \hline
Two-qubit gate execution       & 40                 & $\max(13*\text{num\_ions} - 54, 100)$                       \\ \hline
Segment Traversal             & 40                 & 5                       \\ \hline
Inner-trap swap               & 40                 & 42                       \\ \hline
Trap Split                    & 80                 & 80                        \\ \hline 
Trap Merge                    & 80                 & 80                        \\ \hline
Junction-Y Traversal          & 120                 & 100                       \\ \hline
Junction-X Traversal          & 120                 & 120                       \\ \hline
\end{tabular}
\caption{Benchmarking circuits for the experiments with SHAPER, SHAW, QCCDSim, and 2-stage QCCDSim.}
\label{tab:timming}
\end{table}

\color{teal}
Besides looking into the operation time, which combines the shuttling time and gate execution time. We follow the heating model defined in \cite{murali2020architecting, gutierrez2019transversality} using heating rates of $k_1 = 0.1$ and $k_2 = 0.01$. In general, performing a shuttling operation introduces motion to the system via the trapping potentials, which causes heating. More details about the heating model can be found in \cite{murali2020architecting}. Based on the accumulated heating, the gate fidelity model is defined as follows:
\begin{align}
    \label{eq:gate_fidelity}
    F = 1 - \Gamma \tau - A(2\Bar{n} + 1)
\end{align}
Here, the gate fidelity $F$ is calculated with $\Gamma \coloneq 1$ as the background heating rate of the trap, $\tau$ is the gate execution time as shown in table \ref{tab:timming}, $A \coloneq \max\left(10^{-4}, 10^{-4}\times \frac{\mathrm{num\_ions} }{\ln{\mathrm{num\_ions}}}- 5.3\times10^{-4}\right)$ is a scaling factor on the second term which represents the thermal laser beam instabilities (number of ions, here is the number of ions currently positioned inside the trap), and $\bar{n}$ is the motional mode of the chain \cite{wu2018noise}. When shuttling ions around, the ion gets heated up, where its quantas are then transferred to the motional mode $\bar{n}$ of the trap after merging. After splitting the ion, the introduced quantas is dismissed from the motional mode of the trap (the heating introduced by the split operation still counts). Therefore, the more heat accumulated by the ions along shuttling, the more quanta it introduces to the trap, which decreases the gate fidelity when executing on that trap. Note that this fidelity model is not perfect, and different hardware configurations can have different fidelity models. \blue{To maximize the circuit fidelity, our synthesis objective can be tuned to capture the characteristics of the selected fidelity model.}
\color{black}

\subsection{Experimental results}
\label{sec:exp_result}
We compare our method with the QCCDSim compiler~\cite{murali2020architecting}. We want to emphasize that we did not compare our method with \cite{dai2024advanced} and \cite{zhu2025s} due to the unavailable open-source software. We record two metrics: the final compiled circuit total execution time \color{teal}(total operation time, comprised of shuttling time and gate execution time)\color{black}, as well as the time needed for compilation. We chose QCCDSim due to its capability to derive a near-optimal shuttling schedule for a 2D QCCD architecture. \emph{Moreover, it is the only available framework that works with variable architectures while following the correct description of TI-QCCD architecture.} We evaluated two different QCCD architectures, $H$ and $G2x3$. The $H$ architecture has been pictured previously in Figure~\ref{fig:H_QCCD}. The $G2x3$ architecture is created from the $H$ by adding another column of traps and changing the now-middle Y-junction to an X-junction. In general, the grid architecture is defined as $G(m\times n)$ where $m$ denotes the number of rows and $n$ denotes the number of columns. We vary the maximum ions per trap \color{teal} (\textbf{Trap capacity}) \color{black} and report the \textbf{operation time} in microseconds (\textit{the operation time here is calculated by combining the ion shuttle time and gate execution time.}). QCCDSim is deterministic and only runs once. However, because our initial layout process is random, we run SHAPER and SHAW five times and present the best result. \emph{QCCDSim and 2-stage QCCDSim \textbf{fail} to complete in situations with high congestion; these cases are marked with an} X \emph{in the table.}

\paragraph{Small-scale quantum circuits}
As the mixed-integer linear program (introduced in Appendix \ref{sec:shuttling_MILP}) of the shuttling problem scales poorly (exponential increase in the number of constraints when increasing the number of ions or timestep), we only use it to provide a flavor of the optimal scheduling timing compared to our heuristic for small circuits. In general, SHAPER and SHAW runtimes are less than one minute, while QCCDSim runtime is less than a second. CP-SAT solver for MILP has a timeout of 40,000 seconds. The results for the 8-qubit circuit of SHAPER, SHAW, QCCDSim, and the MILP  are presented in Table~\ref{tab:small-circuit}. In particular, our proposed heuristic manages to \emph{solve all considered configurations while giving significant improvement}, while QCCDSim \emph{fails to resolve half of the cases}.  

In more detail, for these small-scale circuits, on average, SHAPER finds the schedules that are \color{teal} $2.14$ \color{black} times better than QCCDSim, with the best case giving a \color{teal} $3.7$ \color{black} times shorter schedule. SHAW also gives a good performance with schedules that are \color{teal} $1.58$ \color{black} times shorter than QCCDSim and the best case giving \color{teal} $2.66$ \color{black} times better. In all cases, SHAPER and SHAW outperform QCCDSim while needing more time for execution. For MILP, out of 16 configurations, there are 5 cases, we fail to find the optimal schedule, and the runtime already maxes out (longer than $40,000$ seconds). We also observe that solving MILP to find a possible solution for this problem is also extremely hard, and the runtime scales exponentially with the parallelism of the considered circuit. It is also interesting to see that the gap between SHAPER and SHAW, compared to the (sub)optimal MILP, is around \color{teal} $97\%$ \color{black}, while the gap between QCCDSim, compared to the (sub)optimal MILP, is around \color{teal} $285\%$ \color{black}. On average, the runtime of SHAPER is $15.1875$ seconds, the runtime of SHAW is $12.3125$ seconds, the runtime of QCCDSim is $0.4157$ seconds, and the runtime of MILP is $16003.125$ seconds.
\begin{table*}[htbp]
\centering
\begin{tabular}{|c|l|r|r|r|r|r|r|}
\hline

\multicolumn{1}{|c|}{\footnotesize \textbf{Benchmark}} &
\multicolumn{1}{c|}{\footnotesize \textbf{Arch.}} &
\multicolumn{1}{c|}{\footnotesize \textbf{\# trap}} &
\multicolumn{1}{c|}{\footnotesize \color{teal} \textbf{Trap Capacity}  \color{black}} &
\multicolumn{4}{c|}{\footnotesize \textbf{\color{teal} Operation Time} ($\mu s$) \color{black}}\\

\cline{5-8}

&  &  &  &
{\footnotesize \textbf{SHAPER}} & 
{\footnotesize \textbf{SHAW}} & 
{\footnotesize \textbf{QCCDSim}} & 
{\footnotesize \textbf{MILP (CPSAT)}}\\
\hline
\multirow{4}{*}{\footnotesize QAOA\_8} 
& \multirow{2}{*}{G2x2} 
& \multirow{2}{*}{4} 
& 2 
& \textbf{3140} 
& 3720
& X 
& \textit{2240}\\
\cline{4-8}

&  &  & 3 
& \textbf{1800}
& 2400 
& 5760 
& \textit{1600}\\ 
\cline{2-8}

& \multirow{2}{*}{\small G2x3} 
& \multirow{2}{*}{6} 
& 2 
& \textbf{2780} 
& 3000
&  X
& \textit{2120}\\
\cline{4-8}

&  &  & 3 
& \textbf{1880}
& 2640
& 7040
& \textit{1560}\\
\hline

\multirow{4}{*}{\footnotesize QFT\_8} 
& \multirow{2}{*}{G2x2} 
& \multirow{2}{*}{4} 
& 2 
& \textbf{12880}
& 16600
& X
& \textit{7040}\\
\cline{4-8}

&  &  & 3 
& \textbf{10380} 
& 13480
& 17680
& \textit{4440}\\
\cline{2-8}

& \multirow{2}{*}{\small G2x3} 
& \multirow{2}{*}{6} 
& 2 
& \textbf{15440}
& 16720
& X
& \textit{5560}*\\
\cline{4-8}

&  &  & 3 
& \textbf{10440}
& 12560
& 14800
& \textit{5200}*\\
\hline

\multirow{4}{*}{\footnotesize TFIM\_8} 
& \multirow{2}{*}{G2x2} 
& \multirow{2}{*}{4} 
& 2 
& \textbf{5700}
& 6640
& X 
& \textit{7720}\\
\cline{4-8}

&  &  & 3 
& \textbf{3860}
& 5440 
& 6400 
& \textit{1920}\\ 
\cline{2-8}

& \multirow{2}{*}{\small G2x3} 
& \multirow{2}{*}{6} 
& 2
& \textbf{5260}
& 6120
& X 
& \textit{2320}\\
\cline{4-8}

&  &  & 3 
& \textbf{3860}
& 5600 
& 7720 
& \textit{2000} \\
\hline

\multirow{4}{*}{\footnotesize TFXY\_8} 
& \multirow{2}{*}{G2x2} 
& \multirow{2}{*}{4} 
& 2
& \textbf{10660}
& 12640 
& X 
& \textit{4200}* \\
\cline{4-8}

&  &  & 3
& \textbf{8700} 
& 11000 
& 12680 
& \textit{3400}* \\ 
\cline{2-8}

& \multirow{2}{*}{\small G2x3} 
& \multirow{2}{*}{6} 
& 2
& \textbf{11500} 
& 13840 
& X 
& \textit{3800}\\ 
\cline{4-8}

&  &  & 3
& \textbf{8120}
& 10840 
& 15640 
& \textit{3160}* \\
\hline
\end{tabular}
\caption{Comparison among SHAPER, SHAW, QCCDSim, and MILP for various circuits and hardware architectures. Operation time is in microseconds ($\mu s$). \color{teal} For a given architecture, the total number of available ions is calculated by (\# trap) $\times$ Trap capacity. \color{black} 
An ``\textbf{X}'' indicates a failure in resolving the shuttling problem. \textbf{Bold} values in the column(s) indicate the smallest valid shuttle time among those heuristics for that row. \textit{Italic} values in the column(s) indicate the smallest valid shuttle time from solving MILP. The notation $^*$ denotes that MILP fails to find the optimal solution and only returns a sub-optimal solution.}
\label{tab:small-circuit}
\end{table*}

\paragraph{Large-scale quantum circuits}
The results of SHAPER, SHAW, and QCCDSim are presented in Table~\ref{tab:full_result}\color{teal}, and Table~\ref{tab:full-result-large-scale}\color{black}. In particular, our proposed heuristic manages to solve all considered configurations while giving some improvement. QCCDSim fails when the circuit width matches the total trap capacity. These scenarios provide no extra positions in the trap, creating large congestion in the junctions. \emph{In all of these cases, SHAPER and SHAW \textbf{successfully returned} a valid shuttling sequence which is critical for quantum compilation, not to mention its scalability.}

For circuits with sizes from $16$ to $20$ qubits, SHAPER \blue{with 4 passes (a pass consists of one forward pass and one backward pass)} finds compiled sequences on average \blue{$1.936$} times faster \blue{corresponding to $42.5\%$ operation time reduction} when only considering the benchmarks where both algorithms successfully executed. In the best-case trial, we produced a sequence that was \blue{$4$} times faster. The standard deviation of the results from SHAPER is around $4.1\%$. Out of the 32 trials, we produced a sequence that was shorter in \blue{all 32 cases. The use of multiple pass ensure a good initial mapping and better schedule in the end}. SHAPER takes an average of \blue{24.09 seconds} to run across all the benchmarks, while QCCDSim takes \blue{0.055 seconds} on average over those benchmarks that were completed. 
When trying the two-stage approach with QCCDSim with more exhaustive synthesis through the BQSKIT compiler with optimization level $4$ (the best BQSKIT compiler to reduce the gate counts), the result is as expected, where in most cases, the two-stage approach tends to reduce the gate counts but with the trade-off of an increase in either runtime or shuttling cost. In some specific cases, the two-stage approach performs exceptionally well due to the power of PAS/PAM (which only aims to minimize the gate count). For example, Hamiltonian simulation instances (TFIM\_16 and TFXY\_16) have a three-fold reduction in gate counts after performing the synthesis ($450 \rightarrow 150$). We emphasize that SHAPER can also reach this kind of optimality, but with the runtime trade-off.

\begin{figure}
    \centering
    \includegraphics[width=0.85\linewidth]{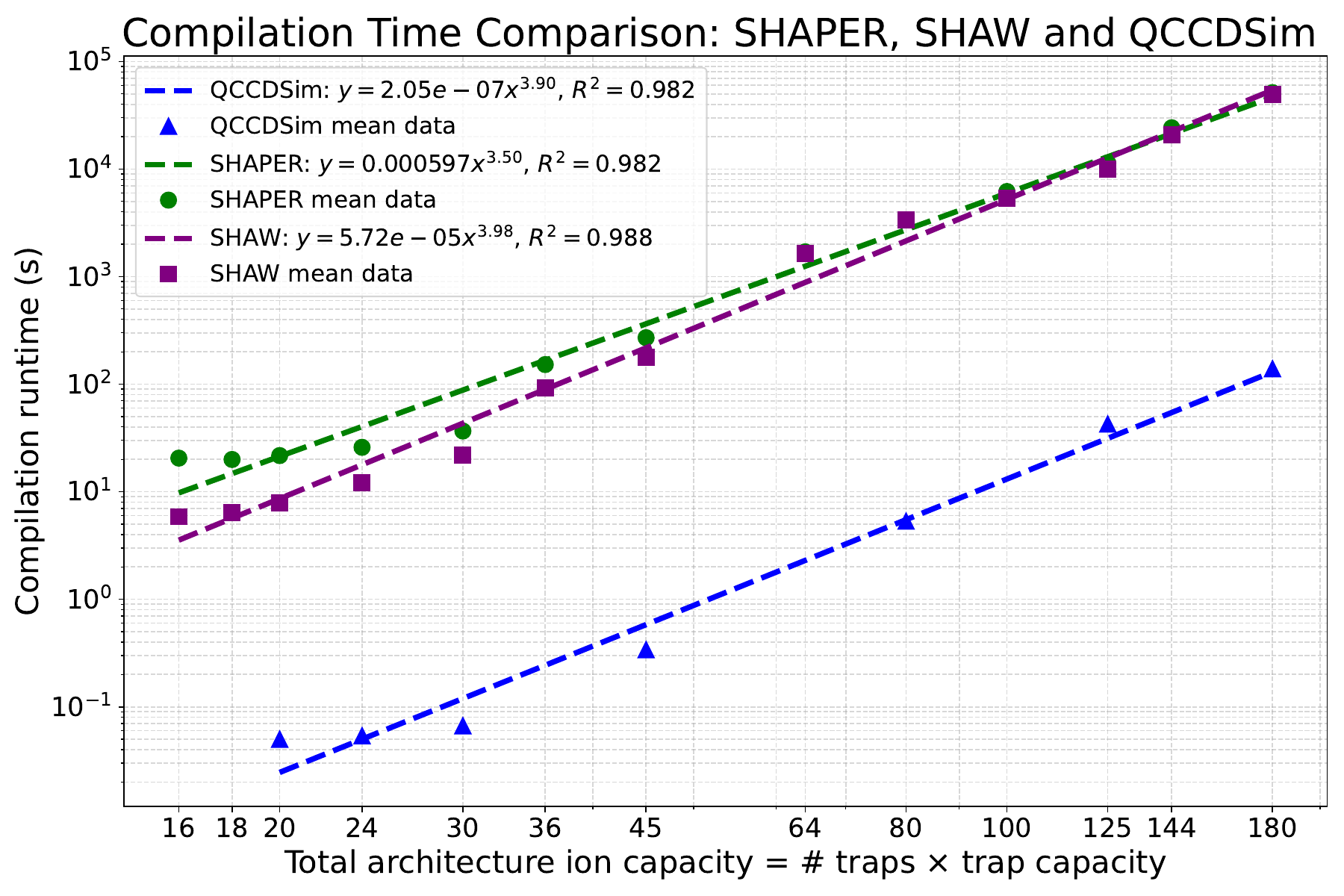}
    \caption{Average compilation runtime (s) plot between SHAPER, SHAW, and QCCDSIM across multiple benchmark circuits at different circuit sizes and architectures. Each data point shows average compile times over different benchmarks at a given architecture (Grid $x \times y$ with $k$ ions per trap).  Here, we consider a power law fitting model for the scaling of compilation runtime, $y = a{x}^{b}$ where $x$ is the total architecture ion capacity. Both SHAPER and SHAW can solve stricter architectures than QCCDSim, as shown in all tables, which results in more data points than QCCDSim. The exponent factor $b$ for both SHAPER and SHAW is around $3.5$ to $3.98$, which is better than QCCDSim, around $3.90$.}
    \label{fig:scaling_fig}
\end{figure}

\color{teal}
For circuits with sizes from $32$ to $128$ qubits, SHAPER with one pass, finds compiled sequences on average $1.68$ times faster when only considering the benchmarks with results for both algorithms. In the best-case trial, we produced a sequence that was $2.21$ times faster. Out of all trials, we always produced a sequence that was shorter. On average, SHAPER and SHAW take more time to run compared to QCCDSim, across those benchmarks that were completed. However, this does not reflect the scaling issues; we see that for larger experiments, QCCDSim runtime scales badly with the number of total ions in an architecture. By performing a power law fitting model $y = a{x}^{b}$, where $x$ is the number of total ions, we can see that the exponent factor $b$ for both SHAPER and SHAW is around $3.5$ to $3.98$, which is better than QCCDSim, around $3.90$ as shown in Figure \ref{fig:scaling_fig}. Although the constant factor $a$ for SHAPER and SHAW is worse than QCCDSim (reflecting why SHAPER and SHAW still take more time than QCCDSim), we emphasize that this can still be improved. 
\blue{Note that an extensive improvement of the compilation runtime can be performed through an engineering step or using a similar scaling method in SABRE, such as using step cost caching like in LightSABRE \cite{zou2024lightsabre} or a multi-level framework as in ML-SABRE \cite{ping2025high}. Here is an example of successful scaling, resulting in LightSHAW introduced in \cite{brent2026scaling}}

\blue{
Another interesting viewpoint is when looking at the utilization of the algorithm when considering a strict architecture. We can observe that QCCDSim fails when the number of qubits matches the number of available trap spaces, or even worse. Let denote the algorithm utilization as $\frac{\text{Circuit Qubits}}{\text{\# Traps}\times \text{Trap Capacity}}\%$ where $100\%$ means the algorithm successfully produces the schedule for the scenario when the qubit circuits match the number of possible ions in the architecture. 
As an example in the $128$-qubit circuits cases, QCCDSim fails in the case of $88\%$ utilization when the total number of ions is $144$ (Grid $6\times6$ architecture with trap capacity of $4$), which is $12.5\%$ more ions than the actual required number of ions. Specifically, across all the benchmarks, QCCDSim only succeeded when the utilization was at most $83.33\%$ and failed for all the cases with higher utilization than $83.33\%$. In contrast, SHAPER can tackle scenarios with $100\%$ utilization, demonstrating the effectiveness of position graph abstraction in deriving an algorithm that produces a competitive shuttling schedule, resolving congestion and deadlock. 
}
\color{black}

As an example to show the potential to fix this scalability problem while maintaining reasonable quality, we compare SHAPER with SHAW (our version of SHAPER but without PAM/PAS). On average, SHAW's is \color{teal} on average $5.3\%$ worse in performance on operation time, but is $2.61$ times faster on $16-20$ qubit circuits and $1.25$ times faster than $32-128$ qubit circuits when compared to SHAPER \color{black}. However, there are specific cases where SHAW actually beats SHAPER, but at the cost of a much bigger runtime. We observed that in these cases, as the routing blocks are not synthesized with respect to the abstracted architecture (making the ions potentially reduce shuttling cost), SHAW has to do much more work to resolve the congestion and find the best scheduling (the gate ordering combined with the architecture constraints creates multiple local minima). We emphasize that this quality-runtime trade-off is desirable because the longer shuttling time results in higher heating, which directly affects the fidelity/robustness of the system \cite{home2009complete, labaziewicz2008suppression}. This is an important aspect, especially with the current quantum hardware state \cite{decross2024computational}.



\begin{table*}[htbp]
\centering
\resizebox{\linewidth}{!}{
\begin{tabular}{|c|l|r|r|r r|r r|r r|r r|}
\hline

\multicolumn{1}{|c|}{\scriptsize \textbf{Benchmark}} &
\multicolumn{1}{c|}{\scriptsize \textbf{Arch.}} &
\multicolumn{1}{c|}{\scriptsize \textbf{\# trap}} &
\multicolumn{1}{c|}{\scriptsize \color{teal} \textbf{Trap Cap} \color{black}} &

\multicolumn{2}{c|}{\footnotesize \textbf{SHAPER}} &                
\multicolumn{2}{c|}{\footnotesize \textbf{SHAW}} &                  
\multicolumn{2}{c|}{\footnotesize \textbf{QCCDSim}} &               
\multicolumn{2}{c|}{\footnotesize \textbf{2-stage QCCDSim}} \\      
\cline{5-12}

&  &  &  &
{\scriptsize \color{teal} \textbf{Operation} \color{black} ($\mu s$)} & 
{\scriptsize \textbf{Runtime} ($s$)} & 
{\scriptsize \color{teal} \textbf{Operation} \color{black} ($\mu s$)} & 
{\scriptsize \textbf{Runtime} ($s$)} & 
{\scriptsize \color{teal} \textbf{Operation} \color{black} ($\mu s$)} & 
{\scriptsize \textbf{Runtime} ($s$)} &
{\scriptsize \color{teal} \textbf{Operation} \color{black} ($\mu s$)} & 
{\scriptsize \textbf{Runtime} ($s$)} \\

\hline
\multirow{4}{*}{\footnotesize QAOA16} 
& \multirow{2}{*}{G2x2} 
& \multirow{2}{*}{4} 
& 4 
& \textbf{\blue{29646}} & \blue{23.03} 
& \blue{31396} & \blue{7.26}
& X & X 
& X & X \\ 
\cline{4-12}

&  &  & 5 
& \blue{23619} & \blue{19.01} 
& \textbf{\blue{23585}} & \blue{8.64}
& 34790 & \blue{0.04} 
& 36688 & 512.75 \\ 
\cline{2-12}

& \multirow{2}{*}{\small G2x3} 
& \multirow{2}{*}{6} 
& 3 
& \textbf{\blue{26293}} & \blue{23.03} 
& \blue{26436} & \blue{6.74}
& X & X 
& X & X \\ 
\cline{4-12}

&  &  & 4 
& \textbf{\blue{20934}} & \blue{21.83} 
& \blue{24588} & \blue{11.28}
& 39063 & \blue{0.06} 
& 37604 & 624.04 \\ 
\hline

\multirow{4}{*}{\footnotesize QV\_16}
& \multirow{2}{*}{G2x2} 
& \multirow{2}{*}{4} 
& 4 
& \textbf{\blue{5639}} & \blue{14.92} 
& \blue{7051} & \blue{3.3}
& X & X 
& X & X \\ 
\cline{4-12}

&  &  & 5 
& \textbf{\blue{2786}} & \blue{13.85} 
& \blue{3549} & \blue{3.93}
& 11162 & \blue{0.03} 
& 14274 & 5311.93 \\ 
\cline{2-12}

& \multirow{2}{*}{\small G2x3} 
& \multirow{2}{*}{6} 
& 3 
& \blue{6069} & \blue{11.49} 
& \textbf{\blue{5677}} & \blue{4.13}
& X & X 
& X & X \\ 
\cline{4-12}

&  &  & 4 
& \textbf{\blue{3498}} & \blue{13.09} 
& \blue{4402} & \blue{4.68}
& 9420 & \blue{0.04} 
& 11277 & 5012.27 \\ 
\hline

\multirow{4}{*}{\footnotesize QFT\_16} 
& \multirow{2}{*}{G2x2} 
& \multirow{2}{*}{4} 
& 4 
& \textbf{\blue{30436}} & \blue{16.05}  
& \blue{33475} & \blue{7.95}
& X & X 
& X & X  \\ 
\cline{4-12}

&  &  & 5 
& \blue{28257} & \blue{19.78} 
& \textbf{\blue{26532}} & \blue{9.5}
& 45488 & \blue{0.06} 
& 44125 & 1155.76 \\ 
\cline{2-12}

& \multirow{2}{*}{\small G2x3} 
& \multirow{2}{*}{6} 
& 3 
& \textbf{\blue{37639}} & \blue{22.72} 
& \blue{43047} & \blue{8.15}
& X & X 
& X & X \\ 
\cline{4-12}

&  &  & 4 
& \blue{33538} & \blue{28.24} 
& \textbf{\blue{32110}} & \blue{13.73}
& 36329 & \blue{0.07} 
& 52787 & 877.52 \\ 
\hline

\multirow{4}{*}{\footnotesize TFIM\_16} 
& \multirow{2}{*}{G2x2} 
& \multirow{2}{*}{4} 
& 4 
& \textbf{\blue{26886}} & \blue{26.57} 
& \blue{30912} & \blue{5.34}
& X & X 
& X & X \\ 
\cline{4-12}

&  &  & 5 
& \textbf{\blue{18012}} & \blue{19.51} 
& \blue{20931} & \blue{7.5}
& 36707 & \blue{0.06} 
& 42722 & 1958.61 \\ 
\cline{2-12}

& \multirow{2}{*}{\small G2x3} 
& \multirow{2}{*}{6} 
& 3 
& \blue{22279} & \blue{24.19} 
& \textbf{\blue{21302}} & \blue{6.20}
& X & X 
& X & X \\ 
\cline{4-12}

&  &  & 4 
& \textbf{\blue{18774}} & \blue{23.29} 
& \blue{19261} & \blue{10.58}
& 42251 & \blue{0.05} 
& 28633 & 1773.41 \\ 
\hline

\multirow{4}{*}{\footnotesize TFXY\_16} 
& \multirow{2}{*}{G2x2} 
& \multirow{2}{*}{4} 
& 4 
& \blue{31413} & \blue{22.12} 
& \textbf{\blue{31149}} & \blue{5.46}
& X & X 
& X & X \\ 
\cline{4-12}

&  &  & 5 
& \textbf{\blue{18971}} & \blue{31.04} 
& \blue{22552} & \blue{7.93}
& 27135 & \blue{0.06} 
& 31356 & 5112.69 \\ 
\cline{2-12}

& \multirow{2}{*}{\small G2x3} 
& \multirow{2}{*}{6} 
& 3 
& \textbf{\blue{26878}} & \blue{18.33} 
& \blue{27953} & \blue{6.67}
& X & X 
& X & X \\ 
\cline{4-12}

&  &  & 4 
& \textbf{\blue{18442}} & \blue{26.67} 
& \blue{20081} & \blue{11.66}
& 45998 & \blue{0.03} 
& 29475 & 4393.78 \\ 
\hline

\multirow{4}{*}{\footnotesize QAOA\_20} 
& \multirow{2}{*}{G2x2} 
& \multirow{2}{*}{4} 
& 5 
& \textbf{\blue{48687}} & \blue{24.74} 
& \blue{53593} & \blue{12.14}
& X & X 
& X & X \\ 
\cline{4-12}

&  &  & 6 
& \textbf{\blue{32785}} & \blue{28.59} 
& \blue{35654} & \blue{18.84}
& 55940 & \blue{0.08} 
& 59045 & 1122.37 \\ 
\cline{2-12}

& \multirow{2}{*}{\small G2x3} 
& \multirow{2}{*}{6} 
& 4 
& \blue{39767} & \blue{25.24} 
& \textbf{\blue{37163}} & \blue{15.26}
& X & X 
& X & X \\ 
\cline{4-12}

&  &  & 5 
& \textbf{\blue{30961}} & \blue{47.24} 
& \blue{39954} & \blue{27.66}
& 51266 & \blue{0.07} 
& 57949 & 1390.36 \\ 
\hline

\multirow{4}{*}{\footnotesize QV\_20} 
& \multirow{2}{*}{G2x2} 
& \multirow{2}{*}{4} 
& 5 
& \textbf{\blue{9527}} & \blue{16.02} 
& \blue{14447} & \blue{5.16}
& X & X 
& X & X \\ 
\cline{4-12}

&  &  & 6 
& \textbf{\blue{6172}} & \blue{22.84} 
& \blue{10474} & \blue{6.70}
& 17674 & \blue{0.02} 
& 16946 & 4407.77 \\ 
\cline{2-12}

& \multirow{2}{*}{\small G2x3} 
& \multirow{2}{*}{6} 
& 4 
& \textbf{\blue{5349}} & \blue{16.80} 
& \blue{5903} & \blue{5.48}
& X & X 
& X & X \\ 
\cline{4-12}

&  &  & 5 
& \blue{4048} & \blue{15.56}
& \textbf{\blue{3960}} & \blue{7.84}
& 15612 & \blue{0.03} 
& 18565 & 4211.96 \\ 
\hline

\multirow{4}{*}{\footnotesize QFT\_20} 
& \multirow{2}{*}{G2x2} 
& \multirow{2}{*}{4} 
& 5 
& \blue{45682} &  \blue{29.53}
& \textbf{\blue{45404}} & \blue{8.58}
& X & X 
& X & X \\ 
\cline{4-12}

&  &  & 6 
& \blue{45938} & \blue{41.53} 
& \textbf{\blue{40683}} & \blue{17.87}
& 56281 & \blue{0.08} 
& 68935 & 1460.02 \\ 
\cline{2-12}

& \multirow{2}{*}{\small G2x3} 
& \multirow{2}{*}{6} 
& 4 
& \textbf{\blue{57632}} & \blue{36.63} 
& \blue{58402} & \blue{18.057}
& X & X 
& X & X \\ 
\cline{4-12}

&  &  & 5 
& \blue{39966} & \blue{47.26}
& \textbf{\blue{39871}} & \blue{30.47}
& 70134 & \blue{0.10}
& 58647 & 1689.41 \\ 
\hline

\end{tabular}
}
\caption{Comparison among SHAPER, SHAW, QCCDSim, and 2-stage QCCDSim for various circuits and hardware 
architectures. 
Shuttle is in microseconds ($\mu s$), runtime in seconds ($s$). Here ``Trap cap'' denotes Trap Capacity. An ``\textbf{X}'' indicates a failure in resolving the shuttling problem when using the corresponding algorithm. 
\textbf{Bold} values indicate the smallest valid shuttle time 
among those heuristic compilers for that row (the considered configuration).}
\label{tab:full_result}
\end{table*}

\color{teal}
\paragraph{Gate count overhead and Circuit fidelity:} As described in Equation \ref{eq:gate_fidelity}, the gate fidelity depends on the accumulated heating of ions. Therefore, the heating introduced by shuttling operations before gate execution dramatically affects the gate fidelity and total circuit fidelity. As stated in Section \ref{sec:exp_setup}, we only consider transpiled circuits for the benchmark. As this transpilation can be exhaustive, the gate count after applying Permutation-aware synthesis has a high chance of staying unchanged. Therefore, our proposed methods, SHAPER and SHAW, do not dramatically modify the gate counts. We want to note that if more weight is added to penalize the congestion objective in the synthesis process described in Section \ref{sec:pam_map}, the gate count can change, but potentially with worse circuit fidelity in general.\\

An interesting observation is that better operation time and better gate counts \emph{do not always guarantee better circuit fidelity for this considered fidelity model unless there is a big enough gap in operation time}. As shown in Table \ref{tab:fidelity-circuit}, when looking at the circuit fidelity of configurations where SHAPER/SHAW and QCCDSim performance have a small gap, for all cases SHAPER/SHAW yield better fidelity. Here, we do not present the cases where the gap is relatively large (for example, in the cases of $8$-qubit circuits and large circuits (from $32$-qubit to $128$-qubit) as in Table \ref{tab:full-result-large-scale}), as the circuit fidelity is guaranteed to be much better compared to QCCDSim. Interestingly, for cases where SHAW and SHAPER have comparable operation times, the schedule with better operation time does not guarantee better circuit fidelity. For example, in the case of TFIM\_16 with the configuration G2x3 and trap capacity of $4$, SHAPER yields a schedule with operation time $18774$, which is better compared to SHAW's schedule ($19261$) but has smaller circuit fidelity, $0.3280$. This is because heating follows ions when merging into the trap and splitting out of the trap, as described when introducing the gate fidelity model in Section \ref{sec:exp_setup}, the circuit fidelity gap depends on the scheduling sequence and circuit structure when considering relatively similar operation time. This suggests another look into the main objective, \emph{where optimizing over operation time may not be sufficient but requires actual hardware information}. 

\color{black}

\begin{table*}[htbp]
\centering
\color{teal}
\resizebox{\linewidth}{!}{
\begin{tabular}{|c|l|c|c|c|c|c|}
\hline

\multicolumn{1}{|c|}{\footnotesize \textbf{Benchmark}} &
\multicolumn{1}{c|}{\footnotesize \textbf{Arch.}} &
\multicolumn{1}{c|}{\footnotesize \textbf{\# trap}} &
\multicolumn{1}{c|}{\footnotesize \textbf{Trap Capacity}} &
\multicolumn{3}{c|}{\footnotesize \textbf{Circuit Fidelity}}\\   
\cline{5-7}
\centering
&  &  &  &
{\footnotesize \textbf{SHAPER}} & 
{\footnotesize \textbf{SHAW}} & 
{\footnotesize \textbf{QCCDSim$^{*}$}}
\\
\hline
\multirow{2}{*}{\footnotesize QAOA\_8} 
& \multirow{1}{*}{G2x2} 
& \multirow{1}{*}{4} 
&3 
& \textbf{0.9227} 
& 0.8465 
& 0.7986 \\ 
\cline{2-7}

& \multirow{1}{*}{\small G2x3} 
& \multirow{1}{*}{6} 
& 3 
& \textbf{0.9220} 
& 0.8495
& 0.7913
\\
\hline

\multirow{2}{*}{\footnotesize QFT\_8} 
& \multirow{1}{*}{G2x2} 
& \multirow{1}{*}{4} 
& 3 
& \textbf{ 0.6704}
& 0.4323
& 0.2942 
\\
\cline{2-7}

& \multirow{1}{*}{\small G2x3} 
& \multirow{1}{*}{6} 
& 3 
& \textbf{0.6420}
& 0.4322
& 0.3456
\\
\hline

\multirow{2}{*}{\footnotesize TFIM\_8} 
& \multirow{1}{*}{G2x2} 
& \multirow{1}{*}{4} 
& 3 
& \textbf{0.8661}
& 0.7410 
& 0.7005 
\\ 
\cline{2-7}

& \multirow{1}{*}{\small G2x3} 
& \multirow{1}{*}{6} 
& 3 
&  \textbf{0.8686}
& 0.7338 

& 0.6999
 \\
\hline

\multirow{2}{*}{\footnotesize TFXY\_8} 
& \multirow{1}{*}{G2x2} 
& \multirow{1}{*}{4} 
& 3
& \textbf{0.7324} 
& 0.5206 
& 0.4360 
\\ 
\cline{2-7}

& \multirow{1}{*}{\small G2x3} 
& \multirow{1}{*}{6} 
& 3
& \textbf{0.7285}
& 0.5280
& 0.4127 
 \\
\hline
\multirow{2}{*}{\footnotesize QFT\_16} 
& \multirow{1}{*}{G2x2} 
& \multirow{1}{*}{4} 
&5
& 0.1565
& \textbf{0.1673}
& \num{5.44e-3}
\\ 
\cline{2-7}

& \multirow{1}{*}{\small G2x3} 
& \multirow{1}{*}{6} 
& 4
&  0.1351
& \textbf{0.1565}
& 0.1281
\\
\hline
\multirow{2}{*}{\footnotesize TFIM\_16} 
& \multirow{1}{*}{G2x2} 
& \multirow{1}{*}{4} 
&5 
& \textbf{0.2869}
& 0.2669 
& 0.1116
\\ 
\cline{2-7}

& \multirow{1}{*}{\small G2x3} 
& \multirow{1}{*}{6} 
& 4
& 0.3280
& \textbf{0.3512}
& 0.1375
\\
\hline
\multirow{2}{*}{\footnotesize TFXY\_16} 
& \multirow{1}{*}{G2x2} 
& \multirow{1}{*}{4} 
& 5 
& \textbf{0.2596}
& 0.2194
& 0.1625
\\ 
\cline{2-7}

& \multirow{1}{*}{\small G2x3} 
& \multirow{1}{*}{6} 
& 4
& \textbf{0.3018}
& 0.2837
& 0.1343
\\
\hline
\multirow{2}{*}{\footnotesize QFT\_20} 
& \multirow{1}{*}{G2x2} 
& \multirow{1}{*}{4} 
& 6
& \num{6.750e-3}
& $\mathbf{7.899 \times 10^{-3}}$
& \num{8.733e-5}
\\ 
\cline{2-7}

& \multirow{1}{*}{\small G2x3} 
& \multirow{1}{*}{6} 
& 5
&  $\mathbf{8.863 \times 10^{-3}}$
& \num{8.620e-3}%
& \num{1.64e-6}
\\
\hline
\end{tabular}
}
\caption{Comparison among SHAPER, SHAW, and QCCDSim$^{*}$ (best schedule from QCCDSim and $2$-stage QCCDSim) for circuits with a small gap of operation times ($\mu$s) from Table \ref{tab:small-circuit} and Table \ref{tab:full_result}. The circuit fidelity has a range from $0 \rightarrow 1$ where circuit fidelity $1$ denotes a noiseless circuit. The gate fidelity model is described in Section \ref{sec:exp_setup}.
\textbf{Bold} values in the column(s) indicate the highest circuit fidelity among those heuristics for that row. }
\label{tab:fidelity-circuit}
\end{table*}

\color{black}

\begin{table*}[htbp]
\centering
\color{teal}
\resizebox{\linewidth}{!}{
\begin{tabular}{|c|l|r|r|r r|r r|r r|}
\hline

\multicolumn{1}{|c|}{\footnotesize \textbf{Benchmark}} &
\multicolumn{1}{c|}{\footnotesize \textbf{Arch.}} &
\multicolumn{1}{c|}{\footnotesize \textbf{\# trap}} &
\multicolumn{1}{c|}{\footnotesize \textbf{Trap Capacity}} &
\multicolumn{2}{c|}{\footnotesize \textbf{SHAPER}} &
\multicolumn{2}{c|}{\footnotesize \textbf{SHAW}} &
\multicolumn{2}{c|}{\footnotesize \textbf{QCCDSim}} \\
\cline{5-10}

&  &  &  &
{\scriptsize \textbf{Operation} ($\mu s$)} &
{\scriptsize \textbf{Runtime} ($s$)} &
{\scriptsize \textbf{Operation} ($\mu s$)} &
{\scriptsize \textbf{Runtime} ($s$)} &
{\scriptsize \textbf{Operation} ($\mu s$)} &
{\scriptsize \textbf{Runtime} ($s$)} \\
\hline

\multirow{2}{*}{\footnotesize QAOA\_32}
& \multirow{2}{*}{G3x3}
& \multirow{2}{*}{9}
& 4
& \textbf{61520} & 194.57
& 65320 & 38.64
& X & X \\
\cline{4-10}

&  &  & 5
& 44120 & 386.19
& \textbf{38871} & 78.6
& 67066  & 0.09 \\
\hline

\multirow{2}{*}{\footnotesize QFT\_32}
& \multirow{2}{*}{G3x3}
& \multirow{2}{*}{9}
& 4
& \textbf{147260} & 118.92
& 179000 & 103.68
& X     & X \\
\cline{4-10}

&  &  & 5
& \textbf{123088} & 236.98
& 129825 & 199.37
& 171663 & 0.49 \\
\hline

\multirow{2}{*}{\footnotesize TFIM\_32}
& \multirow{2}{*}{G3x3}
& \multirow{2}{*}{9}
& 4
& \textbf{154365} & 170.94
& 187100 & 116.73
& X     & X \\
\cline{4-10}

&  &  & 5
& \textbf{138986} & 252.75
& 140186 & 216.09
& 159408 & 0.44 \\
\hline

\multirow{2}{*}{\footnotesize TFXY\_32}
& \multirow{2}{*}{G3x3}
& \multirow{2}{*}{9}
& 4
& 149344 & 154.35
& \textbf{140564} & 110.64
& X     & X \\
\cline{4-10}

&  &  & 5
& 135110 & 235.29
& \textbf{134846} & 214.58
& 143407 & 0.34 \\
\hline

\multirow{2}{*}{\footnotesize QAOA\_64}
& \multirow{2}{*}{G4x4}
& \multirow{2}{*}{16}
& 4
& \textbf{87697} & 376.85
& 97071 & 391.85
& X & X \\
\cline{4-10}

&  &  & 5
& \textbf{82196} & 769.53
& 83504 & 830.31
& 139065 & 0.37\\
\hline

\multirow{2}{*}{\footnotesize QFT\_64}
& \multirow{2}{*}{G4x4}
& \multirow{2}{*}{16}
& 4
& \textbf{288434} & 1114.90
& 300148 & 1031.34 
& X    & X \\
\cline{4-10}

&  &  & 5
& \textbf{274798} & 2551.52
& 278786 & 2072.20
& 569785 & 3.11 \\
\hline

\multirow{2}{*}{\footnotesize TFIM\_64}
& \multirow{2}{*}{G4x4}
& \multirow{2}{*}{16}
& 4
& \textbf{659550} & 2639.98
& 679420 & 2672.33
& X     & X \\
\cline{4-10}

& &  & 5
& \textbf{619295} & 5070.49
& 623649 & 4887.56
& 1085098 & 9.26\\
\hline

\multirow{2}{*}{\footnotesize TFXY\_64}
& \multirow{2}{*}{G4x4}
& \multirow{2}{*}{16}
& 4
& \textbf{650354} & 2667.98
& 664032 & 2482.33
& X     & X \\
\cline{4-10}

&  &  & 5
& 626205 & 5070.26
& \textbf{610140}& 5610.75
& 1003470 & 8.55 \\
\hline
\multirow{2}{*}{\footnotesize QAOA\_98}
& \multirow{2}{*}{G5x5}
& \multirow{2}{*}{25}
& 4
& 234067 & 4665.14
& \textbf{224314} & 3760.96 
& X & X \\
\cline{4-10}

&  &  & 5
& \textbf{212422} & 9707.34
& 227308 & 8041.67
& 396111 & 19.29\\
\hline
\multirow{2}{*}{\footnotesize QFT\_98}
& \multirow{2}{*}{G5x5}
& \multirow{2}{*}{25}
& 4
& \textbf{473492} & 7704.69
& 491855 & 6928.85
& X & X \\
\cline{4-10}

&  &  & 5
& \textbf{439262} & 13095.85 
& 449741 & 11964.86
& 868171 & 65.84 \\
\hline
\multirow{2}{*}{\footnotesize QAOA\_128}
& \multirow{2}{*}{G6x6}
& \multirow{2}{*}{36}
& 4
& \textbf{370420} & 21185.45
& 378367 & 18362.22
& X & X \\
\cline{4-10}

&  &  & 5
& 402398 & 37622.14
& \textbf{356547} & 36337.55
& 795877 & 129.96\\
\hline
\multirow{2}{*}{\footnotesize QFT\_128}
& \multirow{2}{*}{G6x6}
& \multirow{2}{*}{36}
& 4
& 608486 & 27185.66
& \textbf{604953} & 23383.08
& X & X \\
\cline{4-10}

&  &  & 5
& \textbf{602202}  & 65019.91
& 636686 & 62940.00
& 1336085 & 148.00 \\
\hline
\end{tabular}
}
\caption{\color{teal}Large-scale comparison among SHAPER, SHAW, and QCCDSim for various circuits and hardware architectures. Operation time is in microseconds ($\mu s$), runtime in seconds ($s$). An ``\textbf{X}'' indicates a failure in resolving the shuttling problem when using the corresponding algorithm. \textbf{Bold} values indicate the smallest valid operation time among these heuristics for that row.}
\label{tab:full-result-large-scale}
\end{table*}
\color{black}

%% file: isca2024/content/7discussion.tex
\section{Discussion and Conclusion}
\label{sec:discussion}
\paragraph{Better mapping algorithm based on the position graph abstraction}
In this work, we aimed to bridge the gap between the well-studied qubit allocation \cite{siraichi2018qubit} and the routing problem of the superconducting system with the shuttling-based compilation through the position graph abstraction. To prove this point, we introduced the SHuttling-Aware PERmutative search algorithm (SHAPER), which is inspired by the state-of-the-art SABRE \cite{li2019tackling} and the permutation-aware mapping (PAM) \cite{liu2023tackling}. Moreover, as many routing and mapping algorithms share a similar approach with SABRE \cite{liu2022not, niu2020hardware, niu2023enabling, zhu2020dynamic, park2022fast} or based on the established direction such as MAPF \cite{silver2005cooperative, standley2010finding, luna2011push} and deadlock-free routing \cite{anjan1995efficient, domke2011deadlock, dally1987deadlock}, we believe that with elegant and wise adaptation, those methods can be applied to the TI system through our novel position graph with non-trivial adaptation. We believe that, while demonstrating high quality, our heuristic mapping and scheduling algorithms are not optimal, and there exist innovative approaches using the position graph abstraction.

\color{teal} 
\paragraph{Toward fidelity-aware position-graph compilation.}
The fidelity results in this work are evaluated using the model described in Section~\ref{sec:exp_setup}. While this model captures important qualitative
effects of QCCD execution, such as the relationship between shuttling, heating, and gate reliability, it does not fully describe the dynamics of a real trapped-ion QCCD processor. Different fidelity models may emphasize different physical effects and may therefore favor different compilation objectives.

The position graph abstraction naturally enables such model-aware objectives. For example, under the fidelity model used in this work, ions that participate
in many gate operations can accumulate more heating, which lowers the modeled gate fidelity. A synthesis or routing objective could therefore balance the
number of gates acting on each ion, in addition to minimizing gate count, congestion, and shuttling time. This approach can potentially resolve the problem stated in Section \ref{sec:exp_result} “Gate count overhead and Circuit fidelity”, the operation time (shuttling time + gate execution) alone may not be sufficient to capture the circuit fidelity.

However, we do not claim that this model-specific improvement directly implies universal hardware-level fidelity improvement. Without a hardware-calibrated fidelity model, aggressive fidelity optimization may simply overfit to the assumptions of the considering model. A promising direction for future work is therefore to develop calibration-aware or hardware-loop position graph compilation, ideally in collaboration with hardware providers, where the fidelity objective is learned from measured device behavior rather than imposed solely by an abstract model.
\color{black}

\paragraph{Scalability potential of the position graph for QCCD-based TI system}
We argue that the abstraction of the position graph is very scalable and is not only designed for the current state of QCCD-based TI systems. First, the trap zone representation is linear to fit the current state-of-the-art system \cite{sterk2024multi, moses2023race, delaney2024scalable}, however, this can be easily adapted to other types of 2D trap ion arrays in different lattice configurations, such as hexagonal \cite{sterling2014fabrication}, triangular \cite{mielenz2016arrays}, and square \cite{bruzewicz2016scalable}. Moreover, the current state of QCCD systems usually considers Y-junction (T-junction) and X-junction \cite{blakestad2009high, hensinger2006t, bruzewicz2019trapped}(corresponding to $3-$degree and $4-$degree junctions, respectively), while the position graph abstraction can express even higher $d-$degree junctions through the construction of the complete graph. Finally, we argue that the position graph will also be expressible even if the QCCD-based TI system is not planar and becomes a 3D architecture, such as a flip-chip from superconducting qubits \cite{conner2021superconducting}. This is because
We can derive a position graph for each chip layer and put them on
top of each other and represent them through different representation such as hypergraph.

\paragraph{Opportunities with  graph algorithms}
\label{para:use_graph_theory}
The position graph abstraction offers excellent opportunities in terms of applying graph-theoretic methods. Building on the studies of \cite{ratner1990n2, cualinescu2008reconfigurations, yu2015intractability}, numerous heuristic solutions have been developed under specific graph architecture assumptions, such as trees \cite{auletta1999linear}, square grids \cite{cualinescu2008reconfigurations}, and bi-connected graphs \cite{surynek2009novel}. We further propose that adapting well-established algorithms, such as conflict-based search \cite{sharon2015conflict} and increasing cost tree search \cite{sharon2013increasing}, introduces new directions to the field. For instance, utilizing the position graph, we can use node betweenness centrality \cite{freeman1977set} to identify theoretically congested nodes in the studied architectures, where bridges connected to the trap zone are expected congestion points. Moreover, based on \cite{ushijima2019centralities} and the representation of the position graph, we can estimate which ions are likely to exhibit high error rates, as shuttling leads to heating and increased susceptibility to errors. By applying the consumable resources framework from \cite{ushijima2019centralities}, we can model cooling as a consumable resource, allowing us to estimate which ions are prone to excessive heating and, consequently, higher error rates and lower fidelity.

\color{teal}
\paragraph{Potential unifying abstraction across architectures} As we have discussed in Section \ref{sec:position_graph}, our position graph abstraction can be used for both the QCCD-based TI and Superconducting architecture. An interesting future direction is to use the proposed abstraction to generalize to other graph-based quantum architectures. As a concrete example, \cite{kreppel2023quantum, saki2022muzzle} considers the linear trap segment, and multi-qubit gates can only be executed when the ions are positioned into nearest-neighbor configurations. Using our position graph abstraction, we can adapt these constraints by modifying our internal vertices' connections inside the trap (instead of an all-to-all connection, we can use a linear connection). Furthermore, it is an open question whether this abstraction can be used for Neutral Atom systems, especially for the dynamically field-programmable array \cite{tan2024compiling}. 
\color{black}

%% file: isca2024/content/8conclusion.tex

%% file: isca2024/content/9appendix.tex
\section{Mixed-Integer Linear Programming formulation of the Shuttling problem}
\newcommand{\bbf}{\mathbf{b}}
\newcommand{\B}{\mathbf{B}}
\newcommand{\trbf}{\mathbf{tr}}
\newcommand{\TR}{\mathbf{TR}}

\label{sec:shuttling_MILP}
With the definition of the position graph and constraints of QCCD, here we formulate the task of scheduling (routing) the ions on the QCCD architecture using the position graph abstraction to minimize the maximum total processing time over all gates in the quantum circuit. Intuitively,  the MILP formulation of the shuttling problem can be thought of as the combination of the well-known job-shop scheduling problem and MAPF (first, we find the gate (job) that we want to execute (job scheduling), then we try to map the ions to the selected places (MAPF)). To simplify the formulation, the problem is considered in unit time where a move across a given edge costs one unit. The layout of the formulation is based on the position graph abstraction where the nodes represent physical sites and edge representing the transition of qubits (ions) between sites. To represent longer transitions, such as going through a junction, we add dummy sites (i.e., add new nodes in the graph or, in other words, discretize the edge) between edges to make the edge longer. 
\color{teal} As an example, when looking at the timing table in Table \ref{tab:timming}, we first use $40\mu s$ as the cost when ions travel through an edge. Then, for $80\mu s$ split/merge connection requires one dummy node, and the $120\mu s$ junction-$X$ or junction-$Y$ traversal requires two dummy nodes.
\color{black}
Given a quantum circuit with $n$ qubits and DAG mapping $\Pi$ that consists of all native gates, we first define the input data and decision variables as follows
\paragraph{Sets and data}
\begin{itemize}
  \item $I$ – qubits (ions),\quad $P$ – physical sites (nodes of position graph)
  \item $D\subset P$ – \emph{dummy} sites created by long edges (edge $u\to v$ of duration $d>1$ gives $d-1$ dummies nodes $[u \to v\#1, \dots, u \to v\#(d-1)]$). $D_{\rm in}(v)$ denotes dummy sites whose last hop lands on $v$ and $D_{\rm out}(v)$ denotes dummy sites that start a long edge leaving $v$.
  \item $J = \{J_1, J_2 \dots\}, J_i\subset P \times P$ - set of junctions where a junction is presented by edges constructing it connection.
  \item $T=\{0,\dots,H\}$ – discrete ticks,\;
        $E\subset P\times P$ – \emph{unit} edges representing the connections from position graph
        (all non-unit traversals is sliced into $\abs{d}$ unit-length hop via dummy sites).
  \item $\mathcal L$ – undirected \emph{links}  
        ($\ell{=}\min(u,v){-}\max(u,v)$)  
        with capacity $c_\ell\in\{1,2\}$ (2 for “swap” lanes).
  \item $R$ – linear traps; $\;R(p)$ returns the trap containing site $p$.
  \item $\mathcal B$ – two-qubit gates with duration $\mathrm{dur}(b)$,
        qubit set $Q_b$ and precedence relation $(b_1,b_2)\in\Pi$.
  \item $Y_b=\{0,\dots,H-\mathrm{dur}(b)\}$ - is the valid start time of block $b$ where $y_{b,t}$ and $z_{b,r,t}$ are defined 
\end{itemize}

\paragraph{Decision Variables}
\[
    \begin{aligned}
      x_{i,p,t}      &\in\{0,1\} &&\text{ion }i\text{ at site }p\text{ at }t        \\
      f_{i,u,v,t}    &\in\{0,1\} &&\text{ion }i\text{ uses unit edge }(u,v)\text{ at }t\\
      y_{b,t}        &\in\{0,1\} &&\text{gate }b\text{ starts at tick }t \text{ where } t \in Y_b             \\
      z_{b,r,t}      &\in\{0,1\} &&y_{b,t}=1\text{ and trap }r\text{ is chosen}  \\
      C              &\in\mathbb N &&\text{makespan (objective)}
    \end{aligned}
\]
With this, we define the shuttling problem as the problem of minimizing makespan $C$ such that the following group of constraints is satisfied. First, given an initial ion assignment at $t = 0$, we have the following constraint to initialize the ion onto the position graph abstraction.
\begin{align}
     \forall i:  x_{i, \text{init}(i), 0} = 1; \quad \forall p \neq \text{init}(i): x_{i,p,0}=0
\end{align}
Second constraint establish the exclusive occupation of the position graph abstraction by making sure that every qubit(ion) exists at only one place at a given time tick, and each position can only hold at most one qubit(ion).
\begin{align}
    \forall i,p,t:  \sum_{p}x_{i, p, t} = 1, \quad \sum_{i}x_{i, p, t} \leq 1 
\end{align}
We now borrow the celebrated constraint of the network flow problem to create the third constraint that conserves the qubit flow when moving around different positions at unit time, we also add another constraint to make sure ion can only perform one move per tick.
\begin{align}
    &\forall i, p, t  < H:  x_{i, p, t}+\sum_{(u, p)\in E}f_{i, u, p, t} = x_{i, p, t+1}+ \sum_{(p, v)\in E}f_{i, p, v, t} \quad \\
    &\forall i, t < H: \sum_{(u,v)\in E}f_{i, u, v,t} \leq 1
\end{align}
However, this alone is not enough to capture the real hardware constraint of architecture such as QCCD. Therefore, we facilitate it with forth constraint allowing a single move or simultaneous opposite move (“swap”) between different sites. For every edge $\ell$ and tick $t$, with $\ell^{\pm }$ denotes the two different directions and $c_{\ell}$ denotes the edge capacity.
\begin{align}
    \forall \ell, \;t < H:& \nonumber\\ 
     \text{If }c_\ell=1:& \sum_{(u,v) \in \ell^+}\sum_{i}f_{i, u, v, t} + \sum_{(u,v) \in \ell^-}\sum_{i}f_{i, u, v, t} \leq 1  \\
     \text{If }c_\ell=2:& \sum_{(u,v) \in \ell^+}\sum_{i}f_{i, u, v, t} \leq 1, \sum_{(u,v) \in \ell^-}\sum_{i}f_{i, u, v, t} \leq 1  
\end{align}
As stated above, we are considering unit-time flow and use long-edge (with dummy sites) to accommodate it. Therefore, we also introduce the fifth constraint, which (1) forbids an ion to “land on” an already occupied site when finishing a long hop in equation \ref{eq:5_1}, (2) enforces the experimental rule that an ion must idle 1 tick at the target before starting another multi-tick hop in equation \ref{eq:5_2}, and (3) prohibit using dummy sites as parking node which forces ion on a long edge to be traversed continuously without any break on a long edge in equation \ref{eq:5_3} (if ion $i$ in is in dummy node $d$ at tick $t$, ion $i$ need to use unit edge ($d$, successor $d$) to flow at tick $t$).
\begin{align}
\label{eq:5_1}
 &\forall v\in P\setminus D,\;t: \sum_{i}x_{i, v, t}+\sum_{i}\sum_{d\in D_{\text{in}}(v)}x_{i, d, t} \leq 1 \quad  \\
 \label{eq:5_2}
 &\forall v,\;t<H: \sum_{i}\sum_{d\in D_{\text{in}}(v)}f_{i,d,v,t} + \sum_{i}\sum_{d\in D_{\text{out}}(v)}f_{i,v,d,t} \leq 1\\
 \label{eq:5_3}
 &\forall i, d\in D,t<H: x_{i, d,t} = f_{i, d, \text{succ}(d), t}
\end{align}
To finalize the set of constraints for ion flowing, we add the constraints to prohibit multiple ions using one junction. Only one ion is allowed to stay in one junctions at given tick $t$
\begin{align}
    \forall J_i\in J, t < H:\sum_{(u,v) \in J_i}\sum_{i} f_{i,u,v,t}\leq 1,  
\end{align}

Next, we implement the constraint to execute the gate when there exits a trap containing all qubits of that gate. The binary $z_{b,r,t}$ becomes~1 iff \emph{every} qubit of gate~$b$ sits inside trap~$r$ at tick~$t$ and exactly one such trap is selected when the gate starts.
\begin{align}
    & \forall b,r,t,\;i \in Q_b: z_{b,r,t}\le\sum_{p\in r}x_{i, p, t} \\
    &\forall b, t:\sum_{r}z_{b, r, t}=y_{b, t};  \quad \forall b:\sum_{t}y_{b, t}=1 .
\end{align}
Finally, we defines the constraint that ensure the ordering of gate execution based on their dependence from DAG $\Pi$, we also define the variable makespan $C$ where we wish to minimize.
\begin{align}
    &y_{b_1,t_1}+y_{b_2,t_2} \leq 1 \quad \text{when } t_2 < t_1 + \text{dur}(b_1) \text{ for all pairs } (t_1,t_2)\in Y_{b_1}\times Y_{b_2}\\
    &t+\mathrm{dur}(b)\le C \quad \forall b,\;t:y_{b,t}=1.
\end{align}

\color{teal}
\section{Generalization to superconducting qubits}
\label{app:generalization_sc}
The heuristic SHAPER equation \ref{eq:H}, which we derive specifically for QCCD-based TI architecture, can be easily mapped to the well-known heuristic for the SABRE algorithm. As mentioned in section \ref{sec:sc_encoding}, the coupling graph can be easily expressed using a position graph where all positions are executable and measurable, and all edges are swappable and executable. Therefore, 
the term $\sum_{q_i \in B}d(\Phi(q_i))$ in equation \ref{eq:H} is negligible as all positions are executable. Here, the heuristic gets mapped back to the common SABRE heuristic 
\begin{equation}
    \begin{aligned}
        \mathcal{F} &= \frac{1}{\abs{F}} \sum_{B \in F} \sum_{q_i, q_j \in B}  D[\Phi(q_{i})][\Phi(q_{j})] \\
        \mathcal{E} &= \frac{W_{E}}{\abs{E}} \sum_{B \in E}  \sum_{q_i, q_j \in B}  D[\Phi(q_{i})][\Phi(q_{j})] \\
        \mathbf{H} &= \mathcal{F} + \mathcal{E},
    \end{aligned}
\end{equation}
\color{black}

%% file: isca2024/content/rebuttal_refs.bib
@article{brent2026scaling,
  title={Scaling Qubit Mapping and Routing With Position Graph Abstraction and Memoization},
  author={Russon, Brent and Bach, Bao and Younis, Ed and Safro, Ilya },
  journal={arXiv preprint arXiv:2605.09237},
  year={2026}
}


%% file: refs.bib
@article{koch2007charge,
  title={Charge-insensitive qubit design derived from the Cooper pair box},
  author={Koch, Jens and Yu, Terri M and Gambetta, Jay and Houck, Andrew A and Schuster, David I and Majer, Johannes and Blais, Alexandre and Devoret, Michel H and Girvin, Steven M and Schoelkopf, Robert J},
  journal={Physical Review A—Atomic, Molecular, and Optical Physics},
  volume={76},
  number={4},
  pages={042319},
  year={2007},
  publisher={APS}
}

@article{pino2021demonstration,
  title={Demonstration of the trapped-ion quantum CCD computer architecture},
  author={Pino, Juan M and Dreiling, Jennifer M and Figgatt, Caroline and Gaebler, John P and Moses, Steven A and Allman, MS and Baldwin, CH and Foss-Feig, Michael and Hayes, David and Mayer, Karl and others},
  journal={Nature},
  volume={592},
  number={7853},
  pages={209--213},
  year={2021},
  publisher={Nature Publishing Group UK London}
}

@article{herman2023quantum,
  title={Quantum computing for finance},
  author={Herman, Dylan and Googin, Cody and Liu, Xiaoyuan and Sun, Yue and Galda, Alexey and Safro, Ilya and Pistoia, Marco and Alexeev, Yuri},
  journal={Nature Reviews Physics},
  volume={5},
  number={8},
  pages={450--465},
  year={2023},
  publisher={Nature Publishing Group UK London}
}

@Article{cao2019quantum,
  author    = {Cao, Yudong and Romero, Jonathan and Olson, Jonathan P and Degroote, Matthias and Johnson, Peter D and Kieferov{\'a}, M{\'a}ria and Kivlichan, Ian D and Menke, Tim and Peropadre, Borja and Sawaya, Nicolas PD and Sim, Sukin and Veis, Libor and Aspuru-Guzik, Alan},
  title     = {Quantum chemistry in the age of quantum computing},
  number    = {19},
  pages     = {10856--10915},
  volume    = {119},
  journal   = {Chemical reviews},
  publisher = {ACS Publications},
  year      = {2019},
}

@article{biamonte2017quantum,
  title={Quantum machine learning},
  author={Biamonte, Jacob and Wittek, Peter and Pancotti, Nicola and Rebentrost, Patrick and Wiebe, Nathan and Lloyd, Seth},
  journal={Nature},
  volume={549},
  number={7671},
  pages={195--202},
  year={2017},
  publisher={Nature Publishing Group UK London}
}

@article{shaydulin2019hybrid,
  title={A hybrid approach for solving optimization problems on small quantum computers},
  author={Shaydulin, Ruslan and Ushijima-Mwesigwa, Hayato and Negre, Christian FA and Safro, Ilya and Mniszewski, Susan M and Alexeev, Yuri},
  journal={Computer},
  volume={52},
  number={6},
  pages={18--26},
  year={2019},
  publisher={IEEE}
}

@inproceedings{murali2020architecting,
  title={Architecting noisy intermediate-scale trapped ion quantum computers},
  author={Murali, Prakash and Debroy, Dripto M and Brown, Kenneth R and Martonosi, Margaret},
  booktitle={2020 ACM/IEEE 47th Annual International Symposium on Computer Architecture (ISCA)},
  pages={529--542},
  year={2020},
  organization={IEEE}
}

@inproceedings{davis2020towards,
  title={Towards optimal topology aware quantum circuit synthesis},
  author={Davis, Marc G and Smith, Ethan and Tudor, Ana and Sen, Koushik and Siddiqi, Irfan and Iancu, Costin},
  booktitle={2020 IEEE International Conference on Quantum Computing and Engineering (QCE)},
  pages={223--234},
  year={2020},
  organization={IEEE}
}

@inproceedings{younis2021qfast,
  title={Qfast: Conflating search and numerical optimization for scalable quantum circuit synthesis},
  author={Younis, Ed and Sen, Koushik and Yelick, Katherine and Iancu, Costin},
  booktitle={2021 IEEE International Conference on Quantum Computing and Engineering (QCE)},
  pages={232--243},
  year={2021},
  organization={IEEE}
}

@inproceedings{liu2023tackling,
  title={Tackling the qubit mapping problem with permutation-aware synthesis},
  author={Liu, Ji and Younis, Ed and Weiden, Mathias and Hovland, Paul and Kubiatowicz, John and Iancu, Costin},
  booktitle={2023 IEEE International Conference on Quantum Computing and Engineering (QCE)},
  volume={1},
  pages={745--756},
  year={2023},
  organization={IEEE}
}

@inproceedings{li2019tackling,
  title={Tackling the qubit mapping problem for NISQ-era quantum devices},
  author={Li, Gushu and Ding, Yufei and Xie, Yuan},
  booktitle={Proceedings of the twenty-fourth international conference on architectural support for programming languages and operating systems},
  pages={1001--1014},
  year={2019}
}

@article{kielpinski2002architecture,
  title={Architecture for a large-scale ion-trap quantum computer},
  author={Kielpinski, David and Monroe, Chris and Wineland, David J},
  journal={Nature},
  volume={417},
  number={6890},
  pages={709--711},
  year={2002},
  publisher={Nature Publishing Group UK London}
}

@article{moses2023race,
  title={A race-track trapped-ion quantum processor},
  author={Moses, Steven A and Baldwin, Charles H and Allman, Michael S and Ancona, R and Ascarrunz, L and Barnes, C and Bartolotta, J and Bjork, B and Blanchard, P and Bohn, M and others},
  journal={Physical Review X},
  volume={13},
  number={4},
  pages={041052},
  year={2023},
  publisher={APS}
}

@article{mordini2024multi,
  title={Multi-zone trapped-ion qubit control in an integrated photonics QCCD device},
  author={Mordini, Carmelo and Vasquez, Alfredo Ricci and Motohashi, Yuto and M{\"u}ller, Mose and Malinowski, Maciej and Zhang, Chi and Mehta, Karan K and Kienzler, Daniel and Home, Jonathan P},
  journal={arXiv preprint arXiv:2401.18056},
  year={2024}
}

@misc{quantinuum2023roadmap,
  title = {Quantinuum Unveils Accelerated Roadmap to Achieve Universal Fault-Tolerant Quantum Computing by 2030},
  author = {{Quantinuum}},
  year = {2023},
  url = {https://www.quantinuum.com/press-releases/quantinuum-unveils-accelerated-roadmap-to-achieve-universal-fault-tolerant-quantum-computing-by-2030},
  note = {Accessed: 2024-11-12}
}

@inproceedings{siraichi2018qubit,
  title={Qubit allocation},
  author={Siraichi, Marcos Yukio and Santos, Vin{\'\i}cius Fernandes dos and Collange, Caroline and Pereira, Fernando Magno Quint{\~a}o},
  booktitle={Proceedings of the 2018 international symposium on code generation and optimization},
  pages={113--125},
  year={2018}
}

@inproceedings{bhattacharjee2019muqut,
  title={MUQUT: Multi-constraint quantum circuit mapping on NISQ computers},
  author={Bhattacharjee, Debjyoti and Saki, Abdullah Ash and Alam, Mahabubul and Chattopadhyay, Anupam and Ghosh, Swaroop},
  booktitle={2019 IEEE/ACM international conference on computer-aided design (ICCAD)},
  pages={1--7},
  year={2019},
  organization={IEEE}
}

@article{sterk2024multi,
  title={Multi-junction surface ion trap for quantum computing},
  author={Sterk, JD and Blain, MG and Delaney, M and Haltli, R and Heller, E and Holterhoff, AL and Jennings, T and Jimenez, N and Kozhanov, A and Meinelt, Z and others},
  journal={arXiv preprint arXiv:2403.00208},
  year={2024}
}

@article{delaney2024scalable,
  title={Scalable Multispecies Ion Transport in a Grid-Based Surface-Electrode Trap},
  author={Delaney, Robert D and Sletten, Lucas R and Cich, Matthew J and Estey, Brian and Fabrikant, Maya I and Hayes, David and Hoffman, Ian M and Hostetter, James and Langer, Christopher and Moses, Steven A and others},
  journal={Physical Review X},
  volume={14},
  number={4},
  pages={041028},
  year={2024},
  publisher={APS}
}

@article{blakestad2009high,
  title={High-fidelity transport of trapped-ion qubits through an X-junction trap array},
  author={Blakestad, RB and Ospelkaus, C and VanDevender, AP and Amini, JM and Britton, Joseph and Leibfried, Dietrich and Wineland, David J},
  journal={Physical review letters},
  volume={102},
  number={15},
  pages={153002},
  year={2009},
  publisher={APS}
}

@article{hensinger2006t,
  title={T-junction ion trap array for two-dimensional ion shuttling, storage, and manipulation},
  author={Hensinger, WK and Olmschenk, S and Stick, D and Hucul, D and Yeo, M and Acton, Mark and Deslauriers, L and Monroe, C and Rabchuk, J},
  journal={Applied Physics Letters},
  volume={88},
  number={3},
  year={2006},
  publisher={AIP Publishing}
}

@article{bruzewicz2019trapped,
  title={Trapped-ion quantum computing: Progress and challenges},
  author={Bruzewicz, Colin D and Chiaverini, John and McConnell, Robert and Sage, Jeremy M},
  journal={Applied Physics Reviews},
  volume={6},
  number={2},
  year={2019},
  publisher={AIP Publishing}
}

@article{sterling2014fabrication,
  title={Fabrication and operation of a two-dimensional ion-trap lattice on a high-voltage microchip},
  author={Sterling, Robin C and Rattanasonti, Hwanjit and Weidt, Sebastian and Lake, Kim and Srinivasan, Prasanna and Webster, SC and Kraft, Micha{\"e}l and Hensinger, Winfried K},
  journal={Nature communications},
  volume={5},
  number={1},
  pages={3637},
  year={2014},
  publisher={Nature Publishing Group UK London}
}

@article{mielenz2016arrays,
  title={Arrays of individually controlled ions suitable for two-dimensional quantum simulations},
  author={Mielenz, Manuel and Kalis, Henning and Wittemer, Matthias and Hakelberg, Frederick and Warring, Ulrich and Schmied, Roman and Blain, Matthew and Maunz, Peter and Moehring, David L and Leibfried, Dietrich and others},
  journal={Nature communications},
  volume={7},
  number={1},
  pages={ncomms11839},
  year={2016},
  publisher={Nature Publishing Group UK London}
}

@article{bruzewicz2016scalable,
  title={Scalable loading of a two-dimensional trapped-ion array},
  author={Bruzewicz, Colin D and McConnell, Robert and Chiaverini, John and Sage, Jeremy M},
  journal={Nature communications},
  volume={7},
  number={1},
  pages={13005},
  year={2016},
  publisher={Nature Publishing Group UK London}
}

@article{conner2021superconducting,
  title={Superconducting qubits in a flip-chip architecture},
  author={Conner, CR and Bienfait, A and Chang, H-S and Chou, M-H and Dumur, {\'E} and Grebel, J and Peairs, GA and Povey, RG and Yan, H and Zhong, YP and others},
  journal={Applied Physics Letters},
  volume={118},
  number={23},
  year={2021},
  publisher={AIP Publishing}
}

@article{zhu2020exact,
  title={An exact qubit allocation approach for NISQ architectures},
  author={Zhu, Pengcheng and Cheng, Xueyun and Guan, Zhijin},
  journal={Quantum Information Processing},
  volume={19},
  number={11},
  pages={391},
  year={2020},
  publisher={Springer}
}

@article{lin2013ftqls,
  title={FTQLS: Fault-tolerant quantum logic synthesis},
  author={Lin, Chia-Chun and Chakrabarti, Amlan and Jha, Niraj K},
  journal={IEEE Transactions on very large scale integration (VLSI) systems},
  volume={22},
  number={6},
  pages={1350--1363},
  year={2013},
  publisher={IEEE}
}

@article{ge2024quantum,
  title={Quantum circuit synthesis and compilation optimization: Overview and prospects},
  author={Ge, Yan and Wenjie, Wu and Yuheng, Chen and Kaisen, Pan and Xudong, Lu and Zixiang, Zhou and Yuhan, Wang and Ruocheng, Wang and Junchi, Yan},
  journal={arXiv preprint arXiv:2407.00736},
  year={2024}
}

@misc{qiskit2024,
      title={Quantum computing with {Q}iskit},
      author={Javadi-Abhari, Ali and Treinish, Matthew and Krsulich, Kevin and Wood, Christopher J. and Lishman, Jake and Gacon, Julien and Martiel, Simon and Nation, Paul D. and Bishop, Lev S. and Cross, Andrew W. and Johnson, Blake R. and Gambetta, Jay M.},
      year={2024},
      doi={10.48550/arXiv.2405.08810},
      eprint={2405.08810},
      archivePrefix={arXiv},
      primaryClass={quant-ph}
}

@article{ping2025high,
  title={A high-performance multilevel framework for quantum layout synthesis},
  author={Ping, Shuohao and Sathishkumar, Naren and Lin, Wan-Hsuan and Wang, Hanyu and Cong, Jason},
  journal={arXiv preprint arXiv:2505.24169},
  year={2025}
}

@article{zou2024lightsabre,
  title={LightSABRE: A lightweight and enhanced SABRE algorithm},
  author={Zou, Henry and Treinish, Matthew and Hartman, Kevin and Ivrii, Alexander and Lishman, Jake},
  journal={arXiv preprint arXiv:2409.08368},
  year={2024}
}

@inproceedings{murali2019full,
  title={Full-stack, real-system quantum computer studies: Architectural comparisons and design insights},
  author={Murali, Prakash and Linke, Norbert Matthias and Martonosi, Margaret and Abhari, Ali Javadi and Nguyen, Nhung Hong and Alderete, Cinthia Huerta},
  booktitle={Proceedings of the 46th International Symposium on Computer Architecture},
  pages={527--540},
  year={2019}
}

@article{kang2024seeking,
  title={Seeking a quantum advantage with trapped-ion quantum simulations of condensed-phase chemical dynamics},
  author={Kang, Mingyu and Nuomin, Hanggai and Chowdhury, Sutirtha N and Yuly, Jonathon L and Sun, Ke and Whitlow, Jacob and Valdiviezo, Jes{\'u}s and Zhang, Zhendian and Zhang, Peng and Beratan, David N and others},
  journal={Nature Reviews Chemistry},
  pages={1--19},
  year={2024},
  publisher={Nature Publishing Group UK London}
}

@article{foss2024progress,
  title={Progress in trapped-ion quantum simulation},
  author={Foss-Feig, Michael and Pagano, Guido and Potter, Andrew C and Yao, Norman Y},
  journal={Annual Review of Condensed Matter Physics},
  volume={16},
  year={2024},
  publisher={Annual Reviews}
}

@article{schoenberger2024shuttling,
  title={Shuttling for Scalable Trapped-Ion Quantum Computers},
  author={Schoenberger, Daniel and Hillmich, Stefan and Brandl, Matthias and Wille, Robert},
  journal={arXiv preprint arXiv:2402.14065},
  year={2024}
}

@article{schoenberger2024hardware,
  title={Using Compiler Frameworks for the Evaluation of Hardware Design Choices in Trapped-Ion Quantum Computers},
  author={Schoenberger, Daniel and Hillmich, Stefan and Brandl, Matthias and Wille, Robert},
  year={2024}
}

@inproceedings{tseng2024satisfiability,
  title={Satisfiability Modulo Theories-Based Qubit Mapping for Trapped-Ion Quantum Computing Systems},
  author={Tseng, Wei-Hsiang and Chang, Yao-Wen and Jiang, Jie-Hong Roland},
  booktitle={Proceedings of the 2024 International Symposium on Physical Design},
  pages={245--253},
  year={2024}
}

@article{brown2016co,
  title={Co-designing a scalable quantum computer with trapped atomic ions},
  author={Brown, Kenneth R and Kim, Jungsang and Monroe, Christopher},
  journal={npj Quantum Information},
  volume={2},
  number={1},
  pages={1--10},
  year={2016},
  publisher={Nature Publishing Group}
}

@article{ovide2024scaling,
  title={Scaling and assigning resources on ion trap QCCD architectures},
  author={Ovide, Anabel and Cuomo, Daniele and Almudever, Carmen G},
  journal={arXiv preprint arXiv:2408.00225},
  year={2024}
}

@article{durandau2023automated,
  title={Automated generation of shuttling sequences for a linear segmented ion trap quantum computer},
  author={Durandau, Jonathan and Wagner, Janis and Mailhot, Fr{\'e}d{\'e}ric and Brunet, Charles-Antoine and Schmidt-Kaler, Ferdinand and Poschinger, Ulrich and B{\'e}rub{\'e}-Lauzi{\`e}re, Yves},
  journal={Quantum},
  volume={7},
  pages={1175},
  year={2023},
  publisher={Verein zur F{\"o}rderung des Open Access Publizierens in den Quantenwissenschaften}
}

@incollection{dijkstra2022note,
  title={A note on two problems in connexion with graphs},
  author={Dijkstra, Edsger W},
  booktitle={Edsger Wybe Dijkstra: his life, work, and legacy},
  pages={287--290},
  year={2022}
}

@inproceedings{schoenberger2024using,
  title={Using Boolean satisfiability for exact shuttling in trapped-ion quantum computers},
  author={Schoenberger, Daniel and Hillmich, Stefan and Brandl, Matthias and Wille, Robert},
  booktitle={2024 29th Asia and South Pacific Design Automation Conference (ASP-DAC)},
  pages={127--133},
  year={2024},
  organization={IEEE}
}

@inproceedings{saki2022muzzle,
  title={Muzzle the shuttle: efficient compilation for multi-trap trapped-ion quantum computers},
  author={Saki, Abdullah Ash and Topaloglu, Rasit Onur and Ghosh, Swaroop},
  booktitle={2022 Design, Automation \& Test in Europe Conference \& Exhibition (DATE)},
  pages={322--327},
  year={2022},
  organization={IEEE}
}

@article{kreppel2023quantum,
  title={Quantum circuit compiler for a shuttling-based trapped-ion quantum computer},
  author={Kreppel, Fabian and Melzer, Christian and Mill{\'a}n, Diego Olvera and Wagner, Janis and Hilder, Janine and Poschinger, Ulrich and Schmidt-Kaler, Ferdinand and Brinkmann, Andr{\'e}},
  journal={Quantum},
  volume={7},
  pages={1176},
  year={2023},
  publisher={Verein zur F{\"o}rderung des Open Access Publizierens in den Quantenwissenschaften}
}

@article{duong2022quantum,
  title={Quantum neural architecture search with quantum circuits metric and bayesian optimization},
  author={Duong, Trong and Truong, Sang T and Tam, Minh and Bach, Bao and Ryu, Ju-Young and Rhee, June-Koo Kevin},
  journal={arXiv preprint arXiv:2206.14115},
  year={2022}
}

@article{moro2021quantum,
  title={Quantum compiling by deep reinforcement learning},
  author={Moro, Lorenzo and Paris, Matteo GA and Restelli, Marcello and Prati, Enrico},
  journal={Communications Physics},
  volume={4},
  number={1},
  pages={178},
  year={2021},
  publisher={Nature Publishing Group UK London}
}

@article{zhang2020topological,
  title={Topological quantum compiling with reinforcement learning},
  author={Zhang, Yuan-Hang and Zheng, Pei-Lin and Zhang, Yi and Deng, Dong-Ling},
  journal={Physical Review Letters},
  volume={125},
  number={17},
  pages={170501},
  year={2020},
  publisher={APS}
}

@inproceedings{liu2022not,
  title={Not all swaps have the same cost: A case for optimization-aware qubit routing},
  author={Liu, Ji and Li, Peiyi and Zhou, Huiyang},
  booktitle={2022 IEEE International Symposium on High-Performance Computer Architecture (HPCA)},
  pages={709--725},
  year={2022},
  organization={IEEE}
}

@article{niu2020hardware,
  title={A hardware-aware heuristic for the qubit mapping problem in the nisq era},
  author={Niu, Siyuan and Suau, Adrien and Staffelbach, Gabriel and Todri-Sanial, Aida},
  journal={IEEE Transactions on Quantum Engineering},
  volume={1},
  pages={1--14},
  year={2020},
  publisher={IEEE}
}

@article{niu2023enabling,
  title={Enabling multi-programming mechanism for quantum computing in the NISQ era},
  author={Niu, Siyuan and Todri-Sanial, Aida},
  journal={Quantum},
  volume={7},
  pages={925},
  year={2023},
  publisher={Verein zur F{\"o}rderung des Open Access Publizierens in den Quantenwissenschaften}
}

@article{zhu2020dynamic,
  title={A dynamic look-ahead heuristic for the qubit mapping problem of NISQ computers},
  author={Zhu, Pengcheng and Guan, Zhijin and Cheng, Xueyun},
  journal={IEEE Transactions on Computer-Aided Design of Integrated Circuits and Systems},
  volume={39},
  number={12},
  pages={4721--4735},
  year={2020},
  publisher={IEEE}
}

@inproceedings{park2022fast,
  title={A fast and scalable qubit-mapping method for noisy intermediate-scale quantum computers},
  author={Park, Sunghye and Kim, Daeyeon and Kweon, Minhyuk and Sim, Jae-Yoon and Kang, Seokhyeong},
  booktitle={Proceedings of the 59th ACM/IEEE Design Automation Conference},
  pages={13--18},
  year={2022}
}

@inproceedings{younis2022quantum,
  title={Quantum circuit optimization and transpilation via parameterized circuit instantiation},
  author={Younis, Ed and Iancu, Costin},
  booktitle={2022 IEEE International Conference on Quantum Computing and Engineering (QCE)},
  pages={465--475},
  year={2022},
  organization={IEEE}
}

@article{linke2017experimental,
  title={Experimental comparison of two quantum computing architectures},
  author={Linke, Norbert M and Maslov, Dmitri and Roetteler, Martin and Debnath, Shantanu and Figgatt, Caroline and Landsman, Kevin A and Wright, Kenneth and Monroe, Christopher},
  journal={Proceedings of the National Academy of Sciences},
  volume={114},
  number={13},
  pages={3305--3310},
  year={2017},
  publisher={National Acad Sciences}
}

@article{kjaergaard2020superconducting,
  title={Superconducting qubits: Current state of play},
  author={Kjaergaard, Morten and Schwartz, Mollie E and Braum{\"u}ller, Jochen and Krantz, Philip and Wang, Joel I-J and Gustavsson, Simon and Oliver, William D},
  journal={Annual Review of Condensed Matter Physics},
  volume={11},
  number={1},
  pages={369--395},
  year={2020},
  publisher={Annual Reviews}
}

@article{wu2018noise,
  title={Noise analysis for high-fidelity quantum entangling gates in an anharmonic linear paul trap},
  author={Wu, Yukai and Wang, Sheng-Tao and Duan, L-M},
  journal={Physical Review A},
  volume={97},
  number={6},
  pages={062325},
  year={2018},
  publisher={APS}
}

@inproceedings{molavi2022qubit,
  title={Qubit mapping and routing via MaxSAT},
  author={Molavi, Abtin and Xu, Amanda and Diges, Martin and Pick, Lauren and Tannu, Swamit and Albarghouthi, Aws},
  booktitle={2022 55th IEEE/ACM international symposium on Microarchitecture (MICRO)},
  pages={1078--1091},
  year={2022},
  organization={IEEE}
}

@article{malinowski2023wire,
  title={How to wire a 1000-qubit trapped-ion quantum computer},
  author={Malinowski, M and Allcock, DTC and Ballance, CJ},
  journal={PRX Quantum},
  volume={4},
  number={4},
  pages={040313},
  year={2023},
  publisher={APS}
}

@misc{quantinuum_systems, 
author = {Quantinuum}, 
title = {Quantinuum Systems},
year = {2024},
url = {https://www.quantinuum.com/products-solutions/quantinuum-systems}, note = {Accessed: 2024-11-22} }

@article{gutierrez2019transversality,
  title={Transversality and lattice surgery: Exploring realistic routes toward coupled logical qubits with trapped-ion quantum processors},
  author={Guti{\'e}rrez, M and M{\"u}ller, M and Berm{\'u}dez, Alejandro},
  journal={Physical Review A},
  volume={99},
  number={2},
  pages={022330},
  year={2019},
  publisher={APS}
}

@article{gaebler2016high,
  title={High-fidelity universal gate set for be 9+ ion qubits},
  author={Gaebler, John P and Tan, Ting Rei and Lin, Yiheng and Wan, Y and Bowler, Ryan and Keith, Adam C and Glancy, Scott and Coakley, Kevin and Knill, Emanuel and Leibfried, Dietrich and others},
  journal={Physical review letters},
  volume={117},
  number={6},
  pages={060505},
  year={2016},
  publisher={APS}
}

@article{ballance2016high,
  title={High-fidelity quantum logic gates using trapped-ion hyperfine qubits},
  author={Ballance, Christopher J and Harty, Thomas P and Linke, Nobert M and Sepiol, Martin A and Lucas, David M},
  journal={Physical review letters},
  volume={117},
  number={6},
  pages={060504},
  year={2016},
  publisher={APS}
}

@article{farhi2014quantum,
  title={A quantum approximate optimization algorithm},
  author={Farhi, Edward and Goldstone, Jeffrey and Gutmann, Sam},
  journal={arXiv preprint arXiv:1411.4028},
  year={2014}
}

@book{nielsen2001quantum,
  title={Quantum computation and quantum information},
  author={Nielsen, Michael A and Chuang, Isaac L},
  volume={2},
  year={2001},
  publisher={Cambridge university press Cambridge}
}

@article{abrams1999quantum,
  title={Quantum algorithm providing exponential speed increase for finding eigenvalues and eigenvectors},
  author={Abrams, Daniel S and Lloyd, Seth},
  journal={Physical Review Letters},
  volume={83},
  number={24},
  pages={5162},
  year={1999},
  publisher={APS}
}

@article{cross2019validating,
  title={Validating quantum computers using randomized model circuits},
  author={Cross, Andrew W and Bishop, Lev S and Sheldon, Sarah and Nation, Paul D and Gambetta, Jay M},
  journal={Physical Review A},
  volume={100},
  number={3},
  pages={032328},
  year={2019},
  publisher={APS}
}

@techreport{younis2021berkeley,
  title={Berkeley quantum synthesis toolkit (bqskit) v1},
  author={Younis, Ed and Iancu, Costin C and Lavrijsen, Wim and Davis, Marc and Smith, Ethan},
  year={2021},
  institution={Lawrence Berkeley National Laboratory (LBNL), Berkeley, CA (United States)}
}

@book{cormen2009introduction,
  title     = {Introduction to Algorithms},
  author    = {Thomas H. Cormen and Charles E. Leiserson and Ronald L. Rivest and Clifford Stein},
  year      = {2009},
  edition   = {3rd},
  publisher = {MIT Press},
  isbn      = {978-0262033848},
  pages     = {693-700},
  note      = {Section on the Floyd-Warshall algorithm}
}

@article{kokcu2022algebraic,
  title={Algebraic compression of quantum circuits for Hamiltonian evolution},
  author={K{\"o}kc{\"u}, Efekan and Camps, Daan and Bassman Oftelie, Lindsay and Freericks, James K and de Jong, Wibe A and Van Beeumen, Roel and Kemper, Alexander F},
  journal={Physical Review A},
  volume={105},
  number={3},
  pages={032420},
  year={2022},
  publisher={APS}
}

@article{shin2018phonon,
  title={Phonon-driven spin-Floquet magneto-valleytronics in MoS2},
  author={Shin, Dongbin and H{\"u}bener, Hannes and De Giovannini, Umberto and Jin, Hosub and Rubio, Angel and Park, Noejung},
  journal={Nature communications},
  volume={9},
  number={1},
  pages={638},
  year={2018},
  publisher={Nature Publishing Group UK London}
}

@article{decross2024computational,
  title={The computational power of random quantum circuits in arbitrary geometries},
  author={DeCross, Matthew and Haghshenas, Reza and Liu, Minzhao and Rinaldi, Enrico and Gray, Johnnie and Alexeev, Yuri and Baldwin, Charles H and Bartolotta, John P and Bohn, Matthew and Chertkov, Eli and others},
  journal={arXiv preprint arXiv:2406.02501},
  year={2024}
}

@inproceedings{kornhauser1984coordinating,
  title={Coordinating Pebble Motion On Graphs, The Diameter Of Permutation Groups, And Applications},
  author={Kornhauser, D and Miller, G and Spirakis, P},
  booktitle={Proceedings of the 25th Annual Symposium onFoundations of Computer Science, 1984},
  pages={241--250},
  year={1984}
}

@article{cualinescu2008reconfigurations,
  title={Reconfigurations in graphs and grids},
  author={C{\u{a}}linescu, Gruia and Dumitrescu, Adrian and Pach, J{\'a}nos},
  journal={SIAM Journal on Discrete Mathematics},
  volume={22},
  number={1},
  pages={124--138},
  year={2008},
  publisher={SIAM}
}

@article{ratner1990n2,
  title={The (n2- 1)-puzzle and related relocation problems},
  author={Ratner, Daniel and Warmuth, Manfred},
  journal={Journal of Symbolic Computation},
  volume={10},
  number={2},
  pages={111--137},
  year={1990},
  publisher={Elsevier}
}

@inproceedings{papadimitriou1994motion,
  title={Motion planning on a graph},
  author={Papadimitriou, Christos H and Raghavan, Prabhakar and Sudan, Madhu and Tamaki, Hisao},
  booktitle={Proceedings 35th Annual Symposium on Foundations of Computer Science},
  pages={511--520},
  year={1994},
  organization={IEEE}
}

@article{auletta1999linear,
  title={A linear-time algorithm for the feasibility of pebble motion on trees},
  author={Auletta, Vincenzo and Monti, Angelo and Parente, Mimmo and Persiano, Pino},
  journal={Algorithmica},
  volume={23},
  number={3},
  pages={223--245},
  year={1999},
  publisher={Springer}
}

@inproceedings{surynek2009novel,
  title={A novel approach to path planning for multiple robots in bi-connected graphs},
  author={Surynek, Pavel},
  booktitle={2009 IEEE international conference on robotics and automation},
  pages={3613--3619},
  year={2009},
  organization={IEEE}
}

@inproceedings{silver2005cooperative,
  title={Cooperative pathfinding},
  author={Silver, David},
  booktitle={Proceedings of the aaai conference on artificial intelligence and interactive digital entertainment},
  volume={1},
  number={1},
  pages={117--122},
  year={2005}
}

@inproceedings{standley2010finding,
  title={Finding optimal solutions to cooperative pathfinding problems},
  author={Standley, Trevor},
  booktitle={Proceedings of the AAAI conference on artificial intelligence},
  volume={24},
  number={1},
  pages={173--178},
  year={2010}
}

@inproceedings{luna2011push,
  title={Push and swap: Fast cooperative path-finding with completeness guarantees},
  author={Luna, Ryan and Bekris, Kostas E},
  booktitle={IJCAI},
  volume={11},
  pages={294--300},
  year={2011}
}

@inproceedings{ebrahimi2017ebda,
  title={EbDa: A new theory on design and verification of deadlock-free interconnection networks},
  author={Ebrahimi, Masoumeh and Daneshtalab, Masoud},
  booktitle={Proceedings of the 44th Annual International Symposium on Computer Architecture},
  pages={703--715},
  year={2017}
}

@inproceedings{anjan1995efficient,
  title={An efficient, fully adaptive deadlock recovery scheme: DISHA},
  author={Anjan, KV and Pinkston, Timothy Mark},
  booktitle={Proceedings of the 22nd annual international symposium on Computer architecture},
  pages={201--210},
  year={1995}
}

@article{deorio2012reliable,
  title={A reliable routing architecture and algorithm for NoCs},
  author={DeOrio, Andrew and Fick, David and Bertacco, Valeria and Sylvester, Dennis and Blaauw, David and Hu, Jin and Chen, Gregory},
  journal={IEEE Transactions on Computer-Aided Design of Integrated Circuits and Systems},
  volume={31},
  number={5},
  pages={726--739},
  year={2012},
  publisher={IEEE}
}

@inproceedings{domke2011deadlock,
  title={Deadlock-free oblivious routing for arbitrary topologies},
  author={Domke, Jens and Hoefler, Torsten and Nagel, Wolfgang E},
  booktitle={2011 IEEE International Parallel \& Distributed Processing Symposium},
  pages={616--627},
  year={2011},
  organization={IEEE}
}

@article{ogras2006s,
  title={" It's a small world after all": NoC performance optimization via long-range link insertion},
  author={Ogras, Umit Y and Marculescu, Radu},
  journal={IEEE Transactions on very large scale integration (VLSI) systems},
  volume={14},
  number={7},
  pages={693--706},
  year={2006},
  publisher={IEEE}
}

@article{dally1987deadlock,
  title={Deadlock-free message routing in multiprocessor interconnection networks},
  author={Dally and Seitz},
  journal={IEEE Transactions on computers},
  volume={100},
  number={5},
  pages={547--553},
  year={1987},
  publisher={IEEE}
}

@article{van2010graph,
  title={Graph theory and complex networks},
  author={Van Steen, Maarten},
  journal={An introduction},
  volume={144},
  number={1},
  year={2010}
}

@article{blakestad2011near,
  title={Near-ground-state transport of trapped-ion qubits through a multidimensional array},
  author={Blakestad, RB and Ospelkaus, C and VanDevender, AP and Wesenberg, JH and Biercuk, MJ and Leibfried, D and Wineland, David J},
  journal={Physical Review A—Atomic, Molecular, and Optical Physics},
  volume={84},
  number={3},
  pages={032314},
  year={2011},
  publisher={APS}
}

@article{shapira2018robust,
  title={Robust entanglement gates for trapped-ion qubits},
  author={Shapira, Yotam and Shaniv, Ravid and Manovitz, Tom and Akerman, Nitzan and Ozeri, Roee},
  journal={Physical review letters},
  volume={121},
  number={18},
  pages={180502},
  year={2018},
  publisher={APS}
}

@article{weber2024robust,
  title={Robust and fast microwave-driven quantum logic for trapped-ion qubits},
  author={Weber, MA and Gely, MF and Hanley, RK and Harty, TP and Leu, AD and L{\"o}schnauer, CM and Nadlinger, DP and Lucas, DM},
  journal={Physical Review A},
  volume={110},
  number={1},
  pages={L010601},
  year={2024},
  publisher={APS}
}

@article{yu2015intractability,
  title={Intractability of optimal multirobot path planning on planar graphs},
  author={Yu, Jingjin},
  journal={IEEE Robotics and Automation Letters},
  volume={1},
  number={1},
  pages={33--40},
  year={2015},
  publisher={IEEE}
}

@article{sharon2015conflict,
  title={Conflict-based search for optimal multi-agent pathfinding},
  author={Sharon, Guni and Stern, Roni and Felner, Ariel and Sturtevant, Nathan R},
  journal={Artificial intelligence},
  volume={219},
  pages={40--66},
  year={2015},
  publisher={Elsevier}
}

@article{sharon2013increasing,
  title={The increasing cost tree search for optimal multi-agent pathfinding},
  author={Sharon, Guni and Stern, Roni and Goldenberg, Meir and Felner, Ariel},
  journal={Artificial intelligence},
  volume={195},
  pages={470--495},
  year={2013},
  publisher={Elsevier}
}

@article{freeman1977set,
  title={A set of measures of centrality based on betweenness},
  author={Freeman, LC},
  journal={Sociometry},
  year={1977}
}

@article{ushijima2019centralities,
  title={Centralities for networks with consumable resources},
  author={Ushijima-Mwesigwa, Hayato and Khan, Zadid and Chowdhury, Mashrur A and Safro, Ilya},
  journal={Network Science},
  volume={7},
  number={3},
  pages={376--401},
  year={2019},
  publisher={Cambridge University Press}
}

@inproceedings{luo2003real,
  title={Real-time path planning with deadlock avoidance of multiple cleaning robots},
  author={Luo, Chaomin and Yang, Simon X and Stacey, Deborah A},
  booktitle={2003 IEEE International Conference on Robotics and Automation (Cat. No. 03CH37422)},
  volume={3},
  pages={4080--4085},
  year={2003},
  organization={IEEE}
}

@inproceedings{chan2022multi,
  title={Multi-agent pathfinding for deadlock avoidance on rotational movements},
  author={Chan, Frodo Kin Sun and Law, Yan Nei and Lu, Bonny and Chick, Tom and Lai, Edmond Shiao Bun and Ge, Ming},
  booktitle={2022 17th International Conference on Control, Automation, Robotics and Vision (ICARCV)},
  pages={765--770},
  year={2022},
  organization={IEEE}
}

@article{wang2020walk,
  title={Walk, stop, count, and swap: decentralized multi-agent path finding with theoretical guarantees},
  author={Wang, Hanlin and Rubenstein, Michael},
  journal={IEEE Robotics and Automation Letters},
  volume={5},
  number={2},
  pages={1119--1126},
  year={2020},
  publisher={IEEE}
}

@article{home2009complete,
  title={Complete methods set for scalable ion trap quantum information processing},
  author={Home, Jonathan P and Hanneke, David and Jost, John D and Amini, Jason M and Leibfried, Dietrich and Wineland, David J},
  journal={Science},
  volume={325},
  number={5945},
  pages={1227--1230},
  year={2009},
  publisher={American Association for the Advancement of Science}
}

@article{dai2024advanced,
  title={Advanced Shuttle Strategies for Parallel QCCD Architectures},
  author={Dai, Weining and Brown, Kevin A and Robertazzi, Thomas G},
  journal={IEEE Transactions on Quantum Engineering},
  volume={5},
  pages={1--18},
  year={2024},
  publisher={IEEE}
}

@article{labaziewicz2008suppression,
  title={Suppression of heating rates in cryogenic surface-electrode ion traps},
  author={Labaziewicz, Jaroslaw and Ge, Yufei and Antohi, Paul and Leibrandt, David and Brown, Kenneth R and Chuang, Isaac L},
  journal={Physical review letters},
  volume={100},
  number={1},
  pages={013001},
  year={2008},
  publisher={APS}
}

@software{cpsatlp,
  title = {CP-SAT},
  version = { v9.12 },
  author = {Laurent Perron and Frédéric Didier},
  organization = {Google},
  url = {https://developers.google.com/optimization/cp/cp_solver/},
  date = { 2025-02-17 }
}

@inproceedings{zhu2025s,
  title={S-SYNC: Shuttle and Swap Co-Optimization in Quantum Charge-Coupled Devices},
  author={Zhu, Chenghong and Wu, Xian and Wang, Jingbo and Wang, Xin},
  booktitle={Proceedings of the 52nd Annual International Symposium on Computer Architecture},
  pages={271--284},
  year={2025}
}

@article{tan2024compiling,
  title={Compiling quantum circuits for dynamically field-programmable neutral atoms array processors},
  author={Tan, Daniel Bochen and Bluvstein, Dolev and Lukin, Mikhail D and Cong, Jason},
  journal={Quantum},
  volume={8},
  pages={1281},
  year={2024},
  publisher={Verein zur F{\"o}rderung des Open Access Publizierens in den Quantenwissenschaften}
}

@article{pradhan2013finding,
  title={Finding all-pairs shortest path for a large-scale transportation network using parallel Floyd-Warshall and parallel Dijkstra algorithms},
  author={Pradhan, Anu and Mahinthakumar, G},
  journal={Journal of computing in civil engineering},
  volume={27},
  number={3},
  pages={263--273},
  year={2013},
  publisher={American Society of Civil Engineers}
}

@article{leung2018robust,
  title={Robust 2-qubit gates in a linear ion crystal using a frequency-modulated driving force},
  author={Leung, Pak Hong and Landsman, Kevin A and Figgatt, Caroline and Linke, Norbert M and Monroe, Christopher and Brown, Kenneth R},
  journal={Physical review letters},
  volume={120},
  number={2},
  pages={020501},
  year={2018},
  publisher={APS}
}
